\documentclass[%
 aip,
 rsi,
 amsmath,amssymb,
 reprint,%
nofootinbib]{revtex4-2}

\usepackage{graphicx}
\usepackage{dcolumn}
\usepackage{bm}

\usepackage[utf8]{inputenc}
\usepackage[T1]{fontenc}
\usepackage{mathptmx}
\usepackage{etoolbox}
\usepackage[dvipsnames]{xcolor}
\usepackage{footmisc}
\usepackage{comment}

\renewcommand{\vec}{\mathbf}
\renewcommand{\phi}{\varphi}
\newcommand{\SI}{~Supplementary Material}

\makeatother
\begin{document}

{
\makeatletter
\def\frontmatter@thefootnote{%
 \altaffilletter@sw{\@fnsymbol}{\@fnsymbol}{\csname c@\@mpfn\endcsname}%
}%
\makeatother


\title[]{Hierarchical Dynamics and Time-Length Scale Superposition in Glassy Suspensions of Ultra-Low Crosslinked Microgels}

\author{A. Martinelli \textsuperscript{§}}%
\affiliation{L2C, Univ Montpellier, CNRS, F-34095 Montpellier, France}%

\author{R. Elancheliyan \textsuperscript{§}}
\altaffiliation[Present address: ]{Institut Charles Sadron UPR22, Universit\'e de Strasbourg, CNRS, F-67000 Strasbourg, France}
\affiliation{L2C, Univ Montpellier, CNRS, F-34095 Montpellier, France}

\begingroup
    \renewcommand\thefootnote{§}
    \footnotetext{These authors contributed equally.}
\endgroup

\author{A. Scotti}%
\affiliation{Division of Physical Chemistry, Lund University, SE-22100 Lund, Sweden}%

\author{A. V. Petrunin}%
\affiliation{Institute of Physical Chemistry, RWTH Aachen University, Landoltweg 2, 52074 Aachen, Germany}%

\author{D. Truzzolillo}
\email{domenico.truzzolillo@umontpellier.fr}
\affiliation{L2C, Univ Montpellier, CNRS, F-34095 Montpellier, France}

\author{L. Cipelletti}
\email{luca.cipelletti@umontpellier.fr}
\affiliation{L2C, Univ Montpellier, CNRS, F-34095 Montpellier, France}
\affiliation{Institut Universitaire de France, F-75231 Paris, France}
\maketitle
}

\date{\today}
\setcounter{footnote}{0}
\begin{abstract}

We employ small-angle X-ray and dynamic light scattering to investigate the microscopic structure and dynamics of dense suspensions of ultra-low crosslinked (ULC) poly(N-isopropylacrylamide) (PNIPAM) microgels. By probing the supercooled and glassy regimes, we characterize the relationship between structure and dynamics as a function of effective volume fraction $\phi$ and probed length scale. We demonstrate that ULC microgels act as fragile glass formers whose dynamics are governed solely by $\phi$. In contrast, the microscopic structure depends on the specific combination of microgel number density and swelling state that define $\phi$. We identify an anomalous glassy regime where relaxation times are orders of magnitude faster than predicted by supercooled extrapolations, and show that in this regime dynamics are partly accelerated by laser light absorption. Finally, we show that the microscopic relaxation time measured for different $\phi$'s and at various scattering vectors may be rationalized by a ``time-length scale superposition principle'' analogous to the time-temperature superposition used to scale onto a master curve rheology or dielectric relaxation data of molecular systems. Remarkably, we find that the resulting master curve also applies to a different microgel system [V. Nigro \textit{et al.}, Macromolecules \textbf{53}, 1596 (2020)], suggesting a general dynamical behavior of polymeric particles.

\end{abstract}

\maketitle

\section{\label{Intro}Introduction}

Microgels are a prominent class of soft materials, central to contemporary research in colloidal science and engineering. They have distinctive traits, including deformability, permeable nature, and environmental responsiveness \cite{fernandez-nievesMicrogelSuspensionsFundamentals2011,vlassopoulosTunableRheologyDense2014}. Differing significantly on the single-particle level from their hard-sphere counterparts, soft microgels exhibit a highly tunable interaction potential and, consequently, dynamic behavior \cite{vlassopoulosTunableRheologyDense2014,vanderscheerFragilityStrengthNanoparticle2017,nigroRelaxationDynamicsSoftness2020,delmonteTwostepDeswellingVolume2021a,delmonteNumericalStudyNeutral2024,marin-aguilar_unexpected_2025,burger_suspensions_2025}. This makes them ideal candidates for a broad spectrum of advanced applications, ranging from biomedical engineering (such as in drug delivery and tissue scaffold design), to the formulation of smart coatings and the development of advanced sensors, smart materials for oil recovery, and membranes for wastewater treatment~\cite{laiMicrogelsSynthesisProperties2017,fernandez-nievesMicrogelSuspensionsFundamentals2011,uredatReviewStimuliresponsivePolymerbased2024}. 

Microgels have garnered particular attention from the scientific community over the past few decades \cite{vlassopoulosTunableRheologyDense2014,houstonResolvingDifferentBulk2022,vanderscheerFragilityStrengthNanoparticle2017,sethMicromechanicalModelPredict2011}, establishing themselves as highly versatile colloids, whose structure can by tuned both by changing synthesis protocol, chemistry, or via external stimuli\cite{fernandez-nievesMicrogelSuspensionsFundamentals2011}.
They are composed of cross-linked polymer networks swollen in a good solvent. Their unique appeal stems from their inherent "softness" and the remarkable ability to undergo significant changes in their size, shape, and mechanical properties in response to variations in the surrounding environment. This responsiveness is governed by the architecture of their polymeric network and the chemical nature of the monomers used in their synthesis.

In particular, microgels based on poly(N-isopropylacrylamide) (PNIPAM), exhibit remarkable thermosensitivity. 

This property stems from a volume phase transition (VPT), which occurs at a temperature close to the lower critical solution temperature (LCST) of PNIPAM~\cite{peltonParticleSizesElectrophoretic1989}. For PNIPAM in aqueous solution, the LCST is typically around $32^{\circ}$C, a physiologically relevant temperature, making PNIPAM microgels highly attractive for biomedical applications. Below the LCST, PNIPAM chains are hydrophilic due to the formation of hydrogen bonds between the amide groups of the polymer and water molecules, causing the microgels to swell extensively by absorbing a large amount of water. As the temperature is increased above the LCST, these hydrogen bonds weaken, and the polymer chains undergo a coil-to-globule transition. Consequently, water molecules are expelled, leading to a dramatic --yet reversible-- collapse of the microgel network, resulting in a significant reduction in particle size. This thermosensitivity enables dynamic control over microgel properties simply by changing temperature, $T$. The sharpness and magnitude of this transition as well as the colloidal stability of the microgels are furthermore influenced by factors such as crosslink density \cite{ksRevisitingCrosslinkingDensity2025}, copolymerization with other monomers \cite{rodriguezNonuniformSwellingAlkali1994,gujareSmartCopolymerMicrogels2025,elancheliyanRoleChargeContent2022}, and solvent composition \cite{bischofbergerCononsolvencyPNiPAMTransition2014,komarovaBehaviorPNIPAMMicrogels2022}.

The tunable softness of microgels remains one of their most significant physical attributes. Their bulk modulus~\cite{sierra-martinBulkModulusPoly2011} dictates the microgel deformability, interpenetration capabilities, and the bulk rheological behavior of microgel suspensions \cite{conleyJammingOverpackingFuzzy2017,romeoTemperatureControlledTransitionsGlass2010}. The bulk modulus of individual microgels is usually controlled by the density of crosslinks between polymer chains. Conventional microgels are typically synthesized with the addition of a chemical crosslinker, e.g., N,N'-methylenebisacrylamide (BIS) for PNIPAM microgels. The crosslinker molecules form covalent bonds between different polymer chains, establishing the three-dimensional network structure. A higher concentration of crosslinks leads to a denser and more rigid network, resulting in microgels that swell less and exhibit a higher bulk modulus. Accordingly, the choice of the crosslinker, its concentration, and its reactivity significantly impact the internal homogeneity and elasticity of the microgel particles \cite{kyreyInnerStructureDynamics2019,clara-raholaStructuralPropertiesThermoresponsive2012,huControlPolyIsopropylacrylamide2011}.

Ultra-Low Crosslinked (ULC) microgels have recently emerged as an increasingly important class of soft microgels, thanks to their very low bulk modulus~\cite{houstonResolvingDifferentBulk2022}. ULC microgels differ from their conventional counterparts because no crosslinker is added during synthesis. 
In ULC PNIPAM microgels, the formation of a polymeric network relies on the hydrogen atom abstraction mechanism\cite{gaoCrossLinkerFreeIsopropylacrylamideGel2003,gaoInfluenceReactionConditions2003,brugnoniSynthesisStructureDeuterated2019}, leading to self-crosslinking of the PNIPAM chains. During the free radical precipitation polymerization of NIPAM, commonly initiated by species like potassium persulfate (KPS), initiator radicals such as the anion SO$_{4}^{2-}$ issued from the dissociation of KPS, are highly reactive. Not only do  they initiate polymerization by attacking the vinyl bond of the monomer, they may also abstract hydrogen atoms directly from the existing polymer chains, in particular from the tertiary carbon atom within the isopropyl group of the PNIPAM repeating unit. Hydrogen abstraction generates macroradicals --polymer chain radicals-- along the PNIPAM backbone. These macroradicals may recombine with each other or with other propagating radicals, leading to the formation of new carbon-carbon covalent bonds, which effectively act as crosslinks creating a polymer network. Since they involve covalent bonds, these self-crosslinks are as stable as the crosslinks of conventional microgels.

Besides possessing remarkable properties at the single particle level, microgels are also very well-suited for investigating the slowdown of the dynamics in crowded suspensions approaching the colloidal glass transition, which marks the transition from a fluid-like state to a disordered, solid-like state reached at sufficiently high particle volume fraction. Indeed, their versatility makes them ideal model systems for systematically probing the role of softness on this transition. Of particular interest is the possibility of tuning softness, e.g. by varying the crosslink density via the synthesis protocol, and the ability of changing the relevant control parameter, i.e., the effective volume fraction $\phi$ occupied by the microgels, simply by changing $T$, leveraging the swelling/deswelling properties of the microgels. This is in contrast to suspensions of conventional colloidal particles, where $\phi$ may be changed only by preparing distinct samples at different particle number density $n$, a delicate and time-consuming approach that inevitably entails non-negligible uncertainties~\cite{poon_measuring_2012}.

Despite extensive research, however, many questions concerning the structure and dynamics of concentrated microgel suspensions remain open. By analogy with molecular glass formers, colloidal suspensions where the zero-shear viscosity, $\eta$, and the structural relaxation time, $\tau_{\alpha}$, grow exponentially with $\phi$ are dubbed ``strong'', while ``fragile'' suspension exhibit an (apparent) exponential divergence of relaxation times at a finite $\varphi$, described by a colloidal analogous of the  Vogel-Fulcher-Tammann (VFT) law (see also Eq.~\ref{eq:VFT} below). While there is a consensus that colloidal hard spheres act as fragile glass formers, the role of particle softness on fragility is still controversial, with contrasting results reported for microgels~ \cite{mattssonSoftColloidsMake2009,philippe2018glass,gnanMicroscopicRoleDeformation2019,vanderscheerFragilityStrengthNanoparticle2017,nigroRelaxationDynamicsSoftness2020}. A related question, specific to thermosensitive particles, is whether the growth of the microscopic relaxation time with the effective volume fraction follows the same trend when $\phi$ is increased by swelling the particles or by increasing their number density. While this equivalence has been established for the macroscopic viscosity of suspensions of PNIPAM microgels~\cite{sessomsMultipleDynamicRegimes2009}, its microscopic counterpart is still uncharted. At higher $\phi$, in the glassy regime, a surprising flattening of $\tau_{\alpha}(\phi)$ has been reported for several experimental systems~\cite{li_long-term_2017,philippe2018glass,frenzel_glassliquid_2021}, with no equivalent in numerical works, where a steady growth of relaxation times~\cite{delmonteNumericalStudyNeutral2024,marin-aguilar_unexpected_2025} or, on the contrary, a re-entrant melting~\cite{berthierIncreasingDensityMelts2010} have been reported. The origin of this anomalous regime is therefore still to be clarified. Finally, we emphasize that little is known about the structure of dense microgel suspensions~\cite{delmonteNumericalStudyNeutral2024,marin-aguilar_unexpected_2025}, as well as about the length scale behavior of the dynamics of supercooled and glassy suspensions of soft particles~\cite{nigroRelaxationDynamicsSoftness2020,frenzel_glassliquid_2021}.

ULC microgels are particularly suitable to tackle these questions, since they are markedly softer than conventional microgels, yet poorly charged, thus minimizing osmotically-driven self-deswelling, which complicates the interpretation of data  for charged microgels \cite{vanderscheerFragilityStrengthNanoparticle2017}. Furthermore, their very low optical contrast in aqueous solvents, due to their homogeneous internal structure, makes them ideal candidates for light scattering experiments, avoiding complications arising from multiple scattering. 

In this paper, we use small-angle X-ray scattering (SAXS) and state-of-the-art dynamic light scattering to investigate the microscopic structure and dynamics of concentrated ULC microgels suspensions. We show that ULC microgels are fragile glass formers: the microscopic relaxation time increases with $\phi$ following a VFT-like law, with an apparent glass transition volume fraction $\phi_g \sim 1.0 - 1.13$, depending on sample batch. We then demonstrate that the dynamics are ruled solely by $\phi$: samples at the same $\phi$, but different $T$ --and hence swelling state--, or number density $n$, exhibit the same relaxation time. By contrast, we find that the sample structure depends separately on $T$ and $n$. At very high $\phi \gtrsim \phi_g$, ULC microgels exhibit an anomalous regime, where the relaxation time hardly increases with $\phi$ and structural correlations \textit{decrease} with increasing concentration. While both features are reminiscent of the re-entrant melting reported in numerical simulations of soft particles~\cite{berthierIncreasingDensityMelts2010,delmonteNumericalStudyNeutral2024}, here we show that a slight absorption of laser light also plays a role in accelerating the dynamics in this regime. Finally, we show that the dynamics in the deeply supercooled and mildly glassy regimes exhibits a non-trivial scaling of the relaxation time with $q$ vector. Remarkably, we find that data at different $\phi$ fall onto a single $\tau_{\alpha}$ \textit{vs} $q$ master curve when using suitable scaled variables, and that the same master curve describes also recent experiments for a different microgel system~\cite{nigroRelaxationDynamicsSoftness2020}.

The reminder of this paper is organized as follows. In Sec.~\ref{sec:MM} we detail the materials and methods used to synthesize, characterize and probe the structure and dynamics of the suspensions of ULC microgels. In Sec.~\ref{sec:RD} we present and discuss our experimental results.  Finally, in Sec.~\ref{sec:CC} we summarize our key results, followed by some concluding remarks, highlighting the perspectives open by this work.

\section{Materials and methods}\label{sec:MM}

\subsection{Microgel synthesis and characterization}
The synthesis protocol of the ULC microgels is the same as in Refs.~\cite{scotti2020phase,petrunin2023harnessing}.
ULC microgels were obtained from free radical polymerization of NIPAM in absence of a cross-linking agent~\cite{brugnoniSynthesisStructureDeuterated2019}. A reaction solution of 70 mmol/L of NIPAM and 1.2 mmol/L sodium dodecyl sulfate (SDS) in water was purged with nitrogen under stirring at 100 rpm and heated to 70 $^{\circ}$C. At the same time, a solution of potassium peroxydisulfate (KPS) was degassed. The KPS solution was transferred into the monomer solution (final concentration 1.6 mmol/L) to initiate polymerization, which was left to proceed for 4 hours under constant stirring at $70^{\circ}$C. The microgels were purified by three-fold centrifugation at 50000 rpm (relative centrifugal force = 257300). The samples thus obtained were freeze dried for storage.

In this study we used two different batches of ULC microgels, hereafter called ULC1 and ULC2, synthesized following the same aforementioned protocol. For both batches, we determined the proportionality constant $k$ between the concentration $c$ (wt/wt), and the generalized volume fraction $\varphi=kc$ by fitting the zero-shear viscosity $\eta$ of very dilute suspensions ($c\leq$0.004) with the Stokes-Einstein-Batchelor equation~\cite{scotti2020phase}, $\eta/\eta_s=1+2.5kc+5.9(kc)^2$, with $\eta_s$ the viscosity of pure water (see\SI~for details). We obtained two slightly different values: $k_1=44.8\pm 0.9$ and $k_2=48.4\pm 0.01$ for ULC1 and ULC2, respectively. The two batches showed similar, yet not identical, deswelling curves (temperature dependence of the hydrodynamic radius), as shown in Fig. S1 of the\SI. 

The bulk modulus of the ULC1 microgels was determined by measuring deswelling by dynamic light scattering as a function of the concentration of osmotic compressors (linear polyethylene glycol, PEG, of molar mass 35kDa), finding $K=1.43\pm 0.07$~kPa (details in the\SI). A similar characterization was also performed using small-angle neutron scattering with contrast variation, using partially deuterated PEG of molar mass 260 kDa, yielding a comparable value, $K=1.0\pm0.2$~kPa~\cite{houstonResolvingDifferentBulk2022}. These values are about one order of magnitude lower than for standard cross-linked microgels~\cite{philippe2018glass}. Our measurements also showed that $K$ increases with particle compression, up to $\approx 80$ kPa in the tested range of osmotic pressures \cite{houstonResolvingDifferentBulk2022} ($\Pi\leq 159$ kPa).

\subsection{Dynamic Light Scattering}
\label{sec:DLS}
Dynamic light scattering probes the microscopic dynamics through the autocorrelation function of the temporal fluctuations of the scattered intensity~\cite{berne_dynamic_2013}:
\begin{equation}\label{eq:g2-1_DLS}
g_2(\tau,\vec{q})-1 = C \left [\frac{\langle I_{\vec{q}}(t)I_{\vec{q}}(t+\tau)\rangle_t }{\langle I_{\vec{q}}(t)\rangle_t^2 }-1\right] \,,
\end{equation}
where $I_{\vec{q}}(t)$ is the scattered intensity at time $t$ and scattering vector $\vec{q}$, $\langle \cdot\cdot\cdot\rangle_t $ is an average over time, and $C$ is a setup-dependent prefactor such that $g_2(\tau,\vec{q})-1 \rightarrow 1$ for $\tau \rightarrow 0$. The scattering vector depends on the scattering angle $\theta$ at which $I$ is measured: $q = 4\pi n_r \lambda^{-1} \sin \theta/2$, with $n_r$ the solvent refractive index and $\lambda$ the in-vacuo laser wavelength. The intensity correlation function is related to the (coherent, or collective) intermediate scattering function by $F(\tau,\vec{q}) = \sqrt{g_2(\tau,\vec{q})-1}$, where $F(\tau,\vec{q})$ is the Fourier transform of the collective Van Hove probability distribution of collective particle displacements~\cite{berne_dynamic_2013}.  

The experiments reported here were performed using three distinct setups.
For dilute samples, $c<1.78 \times 10^{-2}$, we used a commercial apparatus (Litesizer 500 by Anton Paar), equipped with a red laser ($\lambda=660$~nm). Scattered light was collected at $\theta=90~$deg ($q=17.9~\mu \mathrm{m}^{-1}$). The temperature was set at $T=21.2\pm 0.1$ $^\circ$C by a Peltier element.    
For concentrated suspensions, the majority of the experiments were performed at $\theta=90~$deg using a custom multispeckle~\cite{kirsch1996multispeckle} setup similar to that described in Ref.~\cite{el2009dynamic} (``lab setup'' in the following), with $\lambda= 532.5$~nm ($q=22.2~\mu \mathrm{m}^{-1}$). The samples were placed in a temperature-controlled copper holder and far-field speckles images generated by the scattered light were collected by a CMOS camera (Basler acA2000-340km Camera Link, image size cropped to 640 × 160 pixels). This setup has been used to probe the microscopic dynamics of suspensions where $\varphi$ was changed by varying either temperature $T$ or microgel mass concentration $c$ and hence number density $n$, as detailed later. Finally, a subset of the experiments on a specific concentrated suspension of ULC2 microgels (fixed $c$=0.034, $\varphi$ changed by varying $T$) was performed using COLIS\footnote{Acronym for \textit{Colloidal solids}}, a state-of-the-art light scattering setup developed for experiments onboard the International Space Station (ISS)~\cite{martinelliAdvancedLightScattering2024}. The setup uses laser light with $\lambda= 532.5$~nm  and comprises several measuring lines. Here, we used the small-angle light scattering (SALS) line and three Photon Correlation Imaging lines (PCI45, PCI90, PCI170). The SALS line is equipped with a camera that collects far-field light scattered at angles $0.5~\mathrm{deg} \le \theta \le 9~\mathrm{deg}$. Photon Correlation Imaging is a multispeckle light scattering method were the scattering volume is imaged on a CMOS camera~\cite{duri_resolving_2009}. In COLIS, the PCI lines collect light scattered at $\theta=32.1$, $90$, and $170~$deg ($q=8.7$, $22.2$, and $31.3~\mu \mathrm{m}^{-1}$, respectively). 

For all the concentrated samples ($c>1.78 \times 10^{-2}$) the intensity correlation function $g_2(\tau)$-1 is calculated from a time series of images of the scattered light, acquired with the  variable delay time scheme of Ref.~\cite{philippe_efficient_2016}, to maximize the acquisition efficiency at short time delays. The camera images are analyzed using the multispeckle~\cite{kirsch1996multispeckle} scheme: 
\begin{equation}\label{eq:g2-1_multispeckle}
g_2(\tau)-1 = \langle c_I(t,\tau)\rangle_t \,,
\end{equation}
where the time average is performed on the two-time degree of correlation $c_I$~\cite{duri2005time} :
\begin{equation}\label{eq:cI}
c_I(t,\tau)=C'\left[\frac{\langle I_p(t)I_p(t+\tau)\rangle_p}{\langle I_p(t)\rangle_p \langle I_p(t+\tau)\rangle_p}-1 \right ] \,.
\end{equation}
Here, $I_p(t)$ is the intensity of the $p$-th pixel at time $t$ and $\langle \cdot \cdot \cdot \rangle_p$ indicates an average over camera pixels associated to a small solid angle centered around $\theta$. 
$C'$ is a constant such that $c_I \rightarrow 1$ for $\tau \rightarrow 0$. To reduce the statistical noise due to the finite number of speckles, $C'$ is slightly adjusted for each pair $(t,\tau)$, as detailed in Ref.~\cite{martinelliAdvancedLightScattering2024}. The purpose of the time average of Eq.~\ref{eq:g2-1_multispeckle} is to reduce the experimental noise; it is performed over the full duration of the experiment for stationary samples, or over a short time window of duration $t_{exp}$ where the dynamics are nearly stationary, for samples whose dynamics evolve in time.

\subsection{Small Angle X-Ray Scattering (SAXS)}
The structure of the ULC1 suspensions was investigated by Small Angle X-Ray Scattering (SAXS), performed at the CoSAXS beamline at the 3 GeV ring of the MAX-IV Laboratory (Lund, Sweden)~\cite{plivelicXrayTracingDesign2019}, which is equipped with an Eiger2 4M SAXS detector with pixel size of 75~$\mu$m x 75~$\mu$m. The samples were loaded in quartz capillaries (Hilgenberg) with 1.5 mm diameter and wall thickness of 0.1 mm, placed at a  
sample-to-detector distance of 14.2~m. The X-ray beam energy was $E =12.4~$keV, resulting in $7~\mu\mathrm{m}^{-1} \le q \le 700~\mu\mathrm{m}^{-1}$. A python-based code was used to convert the 2D spectra to 1D $I(q)$ profiles. The data have been corrected for both transmission and background (water in a 1.5 mm capillary).

\section{Results and discussion}\label{sec:RD}

In this work, we focus on concentrated suspensions, approaching or even beyond the glass transition. Accordingly, a key question is to what extent the samples are equilibrated. All samples are initialized by raising $T$ up to $T_{eq}\lesssim 30^{\circ}$C, lower than the volume phase transition temperature of the microgels, $T_c=32.025\pm 0.039^{\circ}$C (see\SI), but high enough to obtain a fully fluid suspension, which, in the COLIS setup, is furthermore homogenized by stirring. Temperature is then lowered to achieve the target $\phi$. We monitor equilibration by inspecting the $t$ dependence of the two-time degree of correlation $c_I$ at fixed time delay $\tau$. As shown in the\SI, in the supercooled regime ($\phi < \phi_g \approx 1.0-1.1$, see below) melting the sample at $T=T_{eq}$ for about 10 min and quenching it at the final $T$ at a rate in the range of $[0.015-0.5]~^{\circ}\mathrm{Cmin}^{-1}$ is sufficient to obtain $t$-independent $c_I$'s, indicative of full equilibration. Above $\phi_g$,  suspensions initialized at $T_{eq}$ are brought to the final $T$ very slowly, with ramps that last up to 48h ($\mathrm{d}T/\mathrm{d}t = 0.0035~^{\circ}\mathrm{Cmin}^{-1}$ and $\mathrm{d}T/\mathrm{d}t = 0.0333~^{\circ}\mathrm{Cmin}^{-1}$ for the lab and COLIS setups, respectively). Once the target $\phi$ is reached and $T$ is kept constant, we find that $c_I$ exhibits a slow growth with $t$, indicative of aging. All measurements reported below for $\phi \geq \phi_g$ are started after letting the sample equilibrate for an additional 48h (lab setup) or 5.6h (COLIS). For these measurements, the degree of correlation is nearly stationary, although it exhibits fluctuations more important than in the supercooled regime (see\SI), as also reported for other microgels~\cite{philippe2018glass}. Overall, we may conclude that samples below $\phi_g$ are well equilibrated, while above $\phi_g$ we are able to reach an almost stationary --yet not fully equilibrated-- state.

\begin{figure*}[t]
\centering
\includegraphics[width=\textwidth]{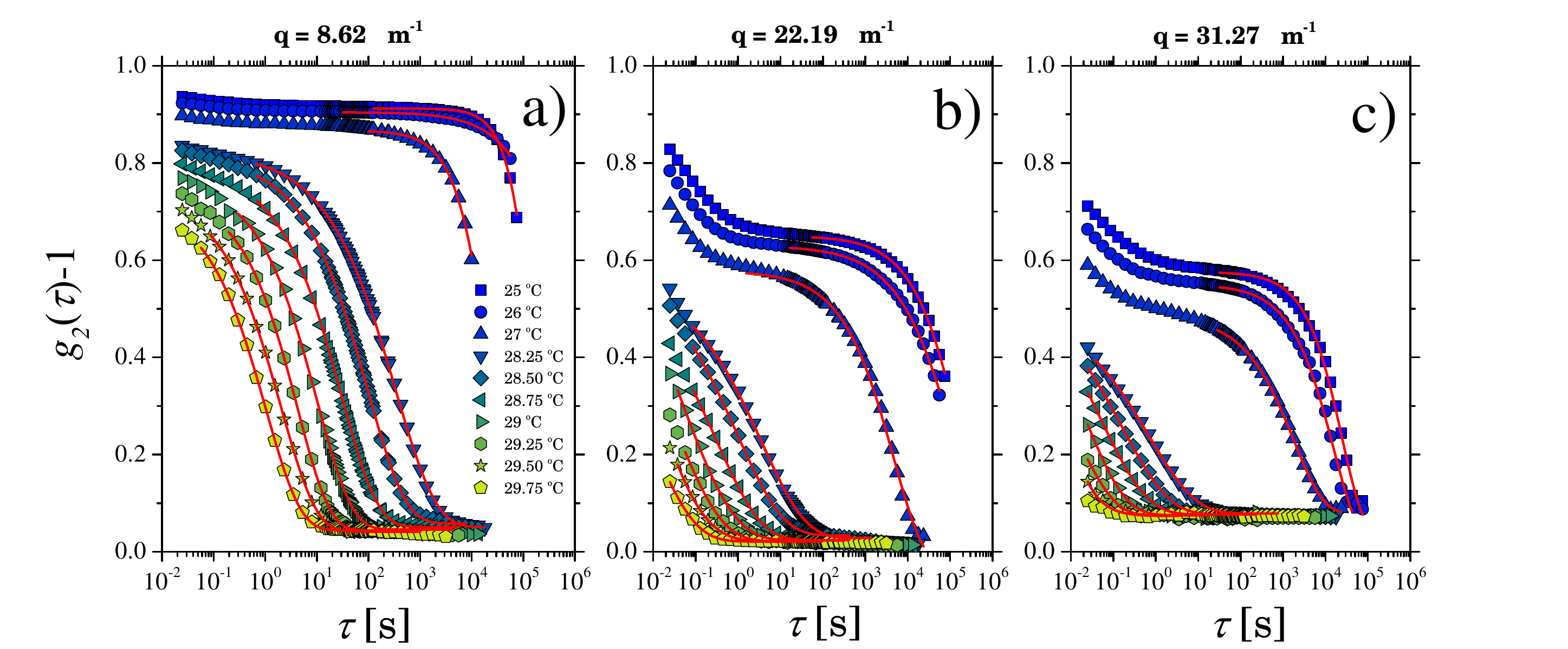}
\caption{\label{fig:wide}Intensity autocorrelation functions $g_2(\tau)-1$ measured in the COLIS setup for an ULC2 sample at fixed $c=0.034$, for various $T$ as shown by the labels. The corresponding $\phi$ increase from 0.875 to 1.29 as $T$ decreases. Panels a)-c) show data acquired simultaneously for three scattering vectors, indicated above each panel. The lines are KWW fits to the data, see Eq.~\ref{eq:KWW}. 
}
\end{figure*}

The evolution of the microscopic dynamics across the transition to glassy behavior is exemplified by representative correlation functions shown in Fig.~\ref{fig:wide}, obtained with the COLIS setup at three distinct scattering vectors, for an ULC2 sample at fixed $c$=0.034 and several $T$'s. The dynamics slow down dramatically upon decreasing $T$ from 29.75~$^{\circ}$C to 25~$^{\circ}$C, corresponding to an increase of $\varphi$ from 0.875 to 1.29. At high $\phi$, $g_2-1$ exhibits a two-step decay, particularly evident at the larger $q$'s. This is a distinctive feature of glassy systems, with the slowest mode corresponding to structural, or $\alpha$, relaxation~\cite{berthier_theoretical_2011}. Note that the fast mode is not fully resolved in our experiments, owing to the limited acquisition rate of the CMOS cameras. In the following, we focus on the slow mode. As seen in Fig.~\ref{fig:wide}, at all $q$'s the slowing down of the dynamics with increasing $\phi$ is accompanied by a change of the shape of the final decay of $g_2-1$, pointing to a qualitative change of the dynamics associated with it.

The relaxation time $\tau_{\alpha}$ and the shape parameter $\beta$ of the $\alpha$-mode are obtained from a fit of the Kohlrausch–Williams–Watts (KWW) function~\cite{berthier_theoretical_2011} to the final decay of the correlation function: 
\begin{equation} \label{eq:KWW}
g_2 (\tau)-1\propto \exp\left [-2(\tau/\tau_{\alpha})^{\beta}\right] + B_{ln} \,,   
\end{equation}
where the factor of two is introduced in the exponential, as in other works~\cite{philippe2018glass,nigroRelaxationDynamicsSoftness2020}, so that $\tau_{\alpha}$ is the relaxation time of the intermediate scattering function, see Sec.~\ref{sec:DLS}. The $B_{ln}$ term accounts for a small but non-zero base line, which in multispeckle experiments typically arises from the non-uniform illumination of the scattering volume~\cite{duri2005time}, with no impact on $\tau_{\alpha}$ and $\beta$.

We start by discussing the $\phi$ dependence of the fitting parameters for measurements at 90~deg ($q=22.1~\mu\mathrm{m}^{-1}$), for which the most comprehensive set of data for both ULC batches and for several temperatures and concentrations is available. Note that varying $T$ entails a change in diffusivity not only because $\phi$ changes, but also because the solvent viscosity $\eta_s$ and the microgel radius $R_h$ are temperature-dependent. We correct for these variations by normalizing $\tau_{\alpha}$ by $\tau_0$, the relaxation time obtained from a KWW fit of $g_2-1$ in the $\phi \rightarrow 0$ regime, where the dynamics are independent of volume fraction. This normalization also allows for including on the same plot low-$\phi$ data collected with the commercial dynamic light scattering setup (Anton Paar), for which the scattering vector is $\approx 20\%$ lower ($q=17.9~\mu\mathrm{m}^{-1}$). 

\begin{figure}[htbp!]
\includegraphics[width=9cm]{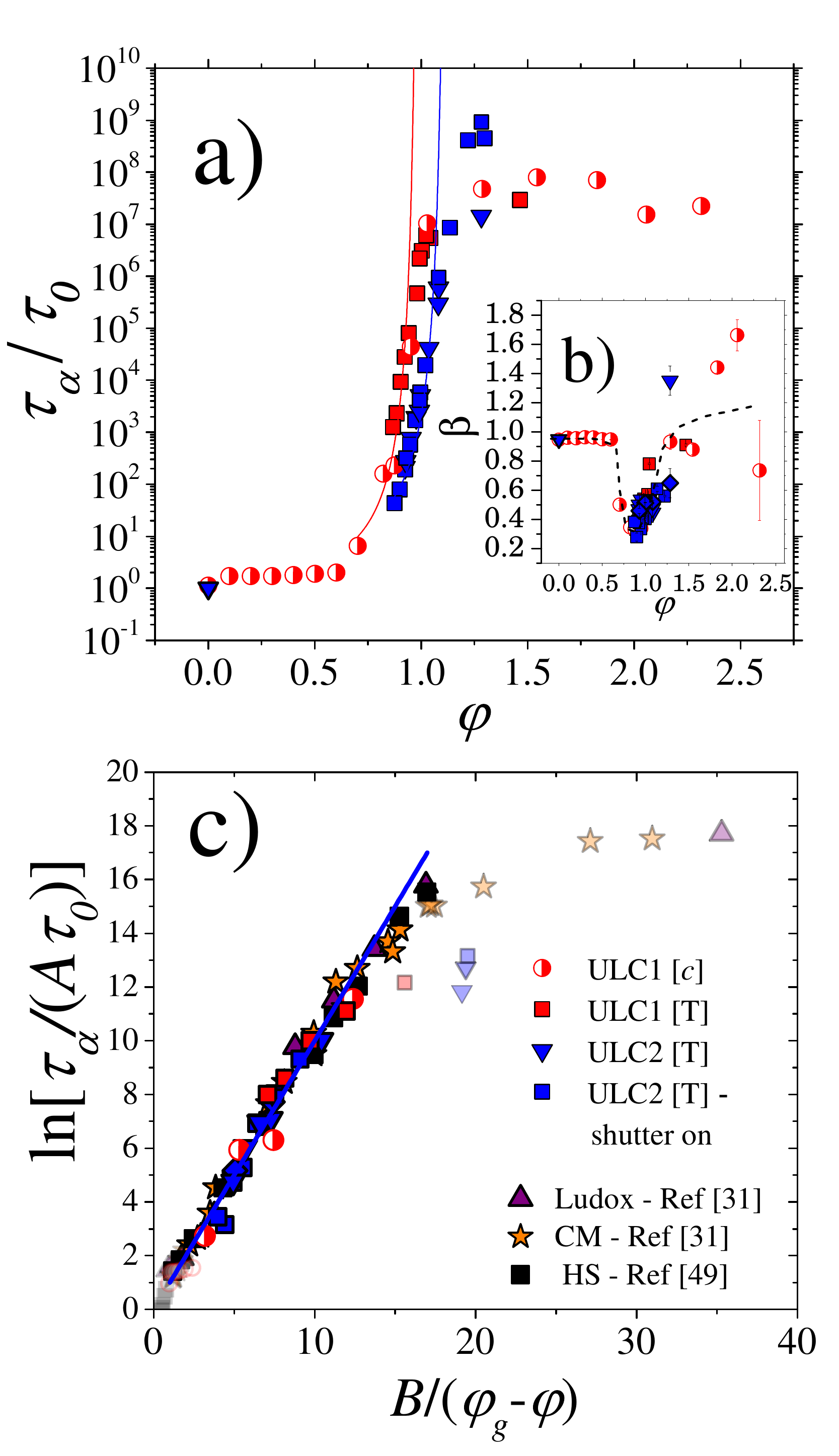}
\caption{\label{fig:tau vs phi} Normalized relaxation time $\tau_{\alpha}/\tau_0$ and stretching exponent $\beta$ (a) and b), respectively) as a function of $\phi$, measured at $q=22.1~\mu \mathrm{m}^{-1}$ for both ULC1 and ULC2. The data labels in c) apply to all panels and indicate how $\phi$ was changed (by varying either $c$ or $T$) and whether a shutter was used to reduce sample illumination. Solid lines in panel a) are VFT fits, Eq.~\ref{eq:VFT}, for data in the range $6.5 \leq \tau_{\alpha}/\tau_0 \leq 28043$ and $42.9 \leq \tau_{\alpha}/\tau_0 \leq 41147$ for ULC1 and ULC2, respectively. In panel c), rescaled variables that linearize the VFT law are used to compare data from this work and for silica nanoparticles (Ludox) and crosslinked microgels (CM)~\cite{philippe2018glass}, and colloidal hard spheres (HS)~\cite{brambilla_probing_2009}. $A$ and $B$ are nearly the same for ULC1 and ULC2 (see text), but differ for the other samples. Semitransparent symbols were not included in the VFT fit. 
}
\end{figure}

Figure~\ref{fig:tau vs phi}a shows that, for both batches (see labels in panel c), $\tau_{\alpha}/\tau_0$ increases by up to nine orders of magnitude upon increasing $\phi$, either by increasing $c$ (semi-filled symbols), or by lowering $T$ (solid symbols). Overall, the behavior seen in Fig.~\ref{fig:tau vs phi}a is similar to that reported for other colloids interacting through a soft potential~\cite{philippe2018glass,nigroRelaxationDynamicsSoftness2020,frenzel_glassliquid_2021}: at low $\phi$, $\tau_{\alpha}$ increases mildly with volume fraction, then it grows rapidly in the supercooled regime (here, approaching $\phi \approx 1$), and finally almost plateaus at very high $\phi$, in the glassy regime. 
These three regimes are also seen in the evolution of the shape parameter: $\beta\approx 1$ at low $\phi$, consistent with the exponential relaxation of $g_2-1$ expected for Brownian particles; in the supercooled regime correlation functions become increasingly stretched ($\beta < 1$, down to values as low as 0.28), while in the glassy regime $\beta$ generally increases above one, although with significant sample-to-sample fluctuations, possibly the result of poorer equilibration at high $\phi$. 

As discussed in Ref.~\cite{philippe2018glass}, the first two regimes are typical of concentrated colloidal suspensions, including hard spheres~\cite{vanmegen_measurement_1998a,brambilla_probing_2009}, while the anomalous $\tau_{\alpha}$ plateau and compressed exponential shape ($\beta>1$) at high $\phi$ are distinctive features of soft colloids.
The scenario emerging for the glass transition of ULC microgels is therefore similar to that for other soft colloids~\cite{philippe2018glass,nigroRelaxationDynamicsSoftness2020,frenzel_glassliquid_2021}. One of the main results in previous works on soft particles, see in particular Refs.~\cite{philippe2018glass,pelaez2015impact, vlassopoulos2001multiarm,vanderscheer_fragility_2017}, was the observation that softness \textit{per se} does not impact fragility, which rather depends on the change of the interparticle potential with $c$, e.g. due to osmotic deswelling. Our ULC microgels are an ideal system to investigate fragility, since they are softer than conventional microgels, but at the same time less prone to osmotic deswelling, owing to the absence of charged groups.

We carefully inspect the supercooled regime where $\tau_{\alpha}$ steeply increases with $\phi$ in Figs.~\ref{fig:tau vs phi}a,c. While the steep slowing down of the dynamics occurs at slightly higher $\phi$ for ULC2 than for ULC1, both batches follow a very similar trend, suggesting that $\tau_{\alpha}(\phi)$ may be fitted using the same functional form. We choose the popular Vogel-Fulcher-Tammann (VFT) form that, for colloidal suspensions, reads
\begin{equation}\label{eq:VFT}
    \tau_{\alpha}(\varphi)=\tau_0 A \exp\left[\frac{B}{\left(\varphi_g-\varphi\right)}\right] \,.
\end{equation}
Here, $\varphi_g$ is the effective volume fraction of the ideal glass transition, where $\tau_{\alpha}$ (apparently) diverges, by extrapolation of the behavior in the supercooled regime. As seen in Fig.~\ref{fig:tau vs phi}a (lines), a very good fit is obtained, yielding the prefactors $A^{(1)}=0.422$ and $A^{(2)}=1.267$  
for ULC1 and ULC2 samples, and $B=0.964$ for both batches. Note that the prefactor of the exponential term is of the same order of magnitude as the microscopic relaxation time in the fully fluid phase, since $A = \mathcal{O}(1)$, an indication that the VFT fit is sound. The glass transition is located at $\varphi_g^{(1)}=1.00 \pm 0.05$ and $\varphi_g^{(2)}=1.13 \pm 0.05$ for ULC1 and ULC2, respectively. We speculate that this difference is due to the different constants $k$ characterizing the two batches, since the relative variation of $k$ and $\varphi_g$ are very similar: $\Delta k/k_1$=0.08 and $\Delta\varphi_g/\varphi_g^{(1)}$=0.13.

To better appreciate the quality of the VFT fit, we re-plot in Fig.~\ref{fig:tau vs phi}c the same data  as in panel a, using scaled variables such that the VFT law corresponds to a straight line through the origin, with slope one (blue line). The two datasets overlap in the entire supercooled regime defined by the fitting range (see caption of Fig.~\ref{fig:tau vs phi}), confirming that similar values of $A$ and the same value of $B$ apply to our ULC microgels, regardless of the batch. Remarkably, an excellent agreement with the VFT law is seen also for other systems with different softness, ranging from hard spheres~\cite{brambilla_probing_2009} to charged-stabilized silica particles~\cite{philippe2018glass}. This clearly confirms that supercooled colloidal suspensions exhibit the same fragile behavior as hard spheres, regardless of their softness, provided that osmotic deswelling is negligible. Note that, for soft particles, $\tau_{\alpha}$  increasingly departs from the VFT law on approaching $\phi_g$, a consequence of the tendency to plateau at large $\phi$ already seen in Fig.~\ref{fig:tau vs phi}a, which we shall discuss later. 

We emphasize that for the ULC microgels varying $\varphi$ by changing particle number density, through $c$, or by changing particle size, through temperature, results in the same change in dynamics, as shown by the red squares and semi-filled circles that fall onto the same curves in Fig.~\ref{fig:tau vs phi}.  Therefore, the change in inter-particle interaction due to the different hydrophobicity of PNIPAM at different $T$ does not affect the microscopic dynamics, a result so far established only for the macroscopic viscosity of conventional , crosslinked PNIPAM microgels~\cite{sessomsMultipleDynamicRegimes2009}. Thus, our data suggest that the slowest relaxation mode of crowded microgel systems is determined only by steric hindrance, governed by $\varphi$.

\begin{figure}[htbp!]
\includegraphics[width=8cm]{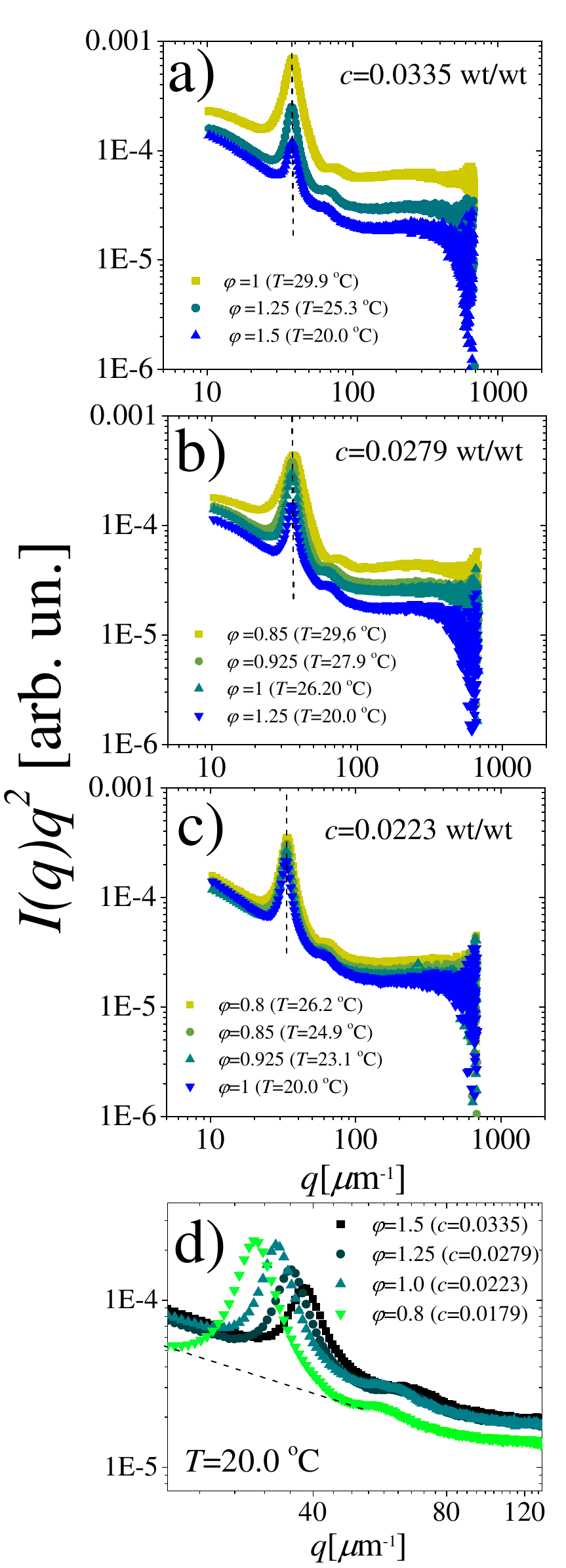}\\
\caption{\label{fig:struct} Kratky plot, $I(q)q^2$ \textit{vs} $q$, for concentrated ULC1 suspensions at three fixed microgel concentrations $c$ and various  $T$ as indicated by the labels (a-c), and at fixed $T=20^{\circ}$C~and various $c$ (d).
}
\end{figure}


We characterize the structure of selected ULC1 samples using small-angle x-ray scattering. In Fig.~\ref{fig:struct} we show Kratky plots, $I(q)q^2$ \textit{vs} $q$, for three different concentrations $c$ and various temperatures, in a range of $\phi$ that encompasses the glass transition, up to the $\phi>\phi_g$ regime where the relaxation time saturates. We choose the Kratky representation rather than calculating the structure factor $S(q)$ since structural features, such as a peak signalling interparticle position correlations, clearly emerge in a Kratky plot, with no need of normalizing $I(q)$ by the particle form factor $P(q)$ to obtain $S(q)$, as it is usually done in scattering experiments~\cite{cipelletti_static_2024a}. Indeed, $P(q)$ is difficult to determine reliably for soft, deformable particles. 
All spectra show a clear main diffraction peak, whose position at fixed $c$ does not depend on $T$, Fig.~\ref{fig:struct}a-c, while at fixed $T$ it shows a marked dependence on microgel number density (or, equivalently, $c$), as shown in Fig. \ref{fig:struct}d.

\begin{figure}[htbp!]
\includegraphics[width=10cm]{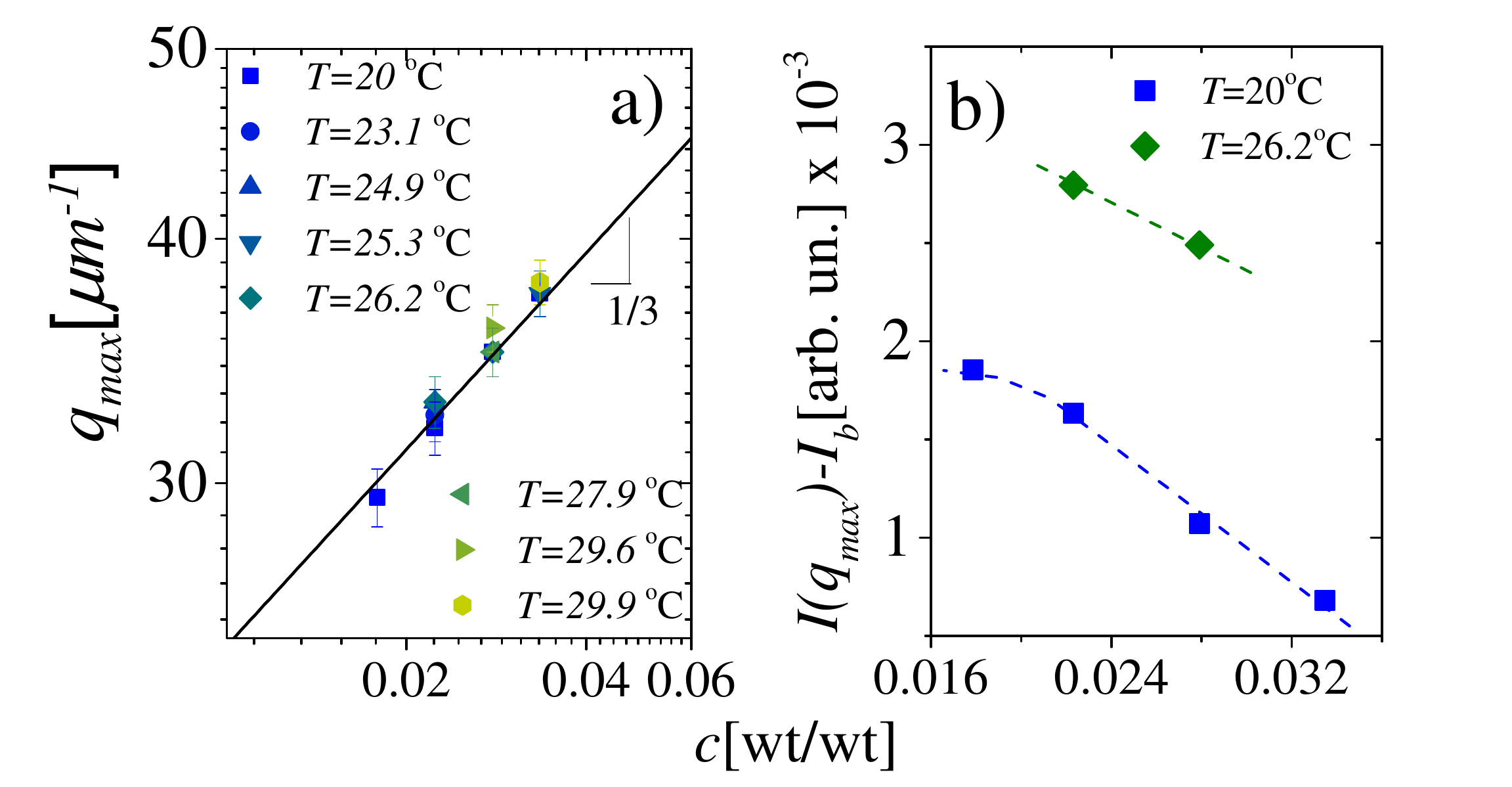}\\
\caption{\label{fig:struct2} a) Peak position $q_{max}$ and b) height of the first peak in function of microgel concentration c[wt/wt] and 2 different temperatures as indicated in the panel}
\end{figure}

To carefully characterize the position and height of the peak, we subtract from $I(q)q^2$ a baseline obtained by prolonging the power-law behavior $I(q)q^2 \propto q^{-\alpha}$, with $\alpha\approx 0.5$--$1$, measured on the left side of each peak (see, e.g., the dashed line in Fig.~\ref{fig:struct}d). The position (all data) and height (samples at constant $T$) of the peaks after baseline subtraction are reported in Fig.~\ref{fig:struct2}. We first discuss the peak position. Figure~\ref{fig:struct2}a confirms that, at fixed $c$, the peak position $q_{max}$ does not vary with $T$, to within experimental uncertainty. By contrast, $q_{max}$ scales as $c^{1/3}$, regardless of $T$. This behavior is consistent with the geometric scaling of $d_c$, the average interparticle distance, as a function of  concentration expected for dense suspensions~\cite{cipelletti_static_2024a}, where $d_c\sim 1/q_{max}\sim c^{-1/3}$. While this scaling is not surprising, we remind that the dynamics of all our ULC samples follows a single VFT law in the supercooled regime upon varying either $c$ or $T$ (Fig.~\ref{fig:tau vs phi}). We are thus left with samples whose dynamics vary in the same way with $\phi$, regardless of the control parameter ($c$ ot $T$), while their structure evolves very differently, a highly non-trivial result, which will need further theoretical modelling to be fully understood.

The $c$ dependence of the baseline-corrected peak height is shown in Fig.~\ref{fig:struct2}, for two data sets collected at $T=20~^{\circ}$C and $26.2~^{\circ}$C. These data correspond to volume fractions in the range $\phi = 0.8$--$1.5$, from slightly below $\phi_g = 1$ to well within the glassy regime. The general trend is for the peak height to \textit{decrease} with increasing $\phi$ (i.e. increasing $c$ at fixed $T$ or decreasing $T$ at a given $c$, see the offset between the two data series). This trend is surprising, since an overall growth of ordering is expected in dense colloidal suspensions when $\phi$ grows, resulting in higher structural peaks~\cite{TheorySimpleLiquids2013}. Note that it is very unlikely the decrease of the peak height results from a change of $P(q)$: in the very dense regime probed here, the microgels are in contact, such that the characteristic features of $P(q)$ and $S(q)$ (roll-off at high $q$ and a peak, respectively) scale in the same way with $d_c$ and hence with $c$. 
Instead, the decrease of the peak amplitude reflects a tendency towards structural melting across the glass transition, as previously reported for crosslinked microgels\cite{philippe2018glass,zhangThermalVestigeZerotemperature2009}, charged silica particles\cite{philippe2018glass}, and nanoemulsions \cite{gravesStructureConcentratedNanoemulsions2005}, and also observed in numerical simulations of both model soft particles~\cite{berthierIncreasingDensityMelts2010} and more realistic conventional microgels~\cite{delmonteNumericalStudyNeutral2024}, although, surprisingly, not for simulated ULC microgels~\cite{marin-aguilar_unexpected_2025}. This non-trivial effect has been ascribed to the fact that, for very soft and highly concentrated systems, the entropy gain associated with the exploration of a large number of disordered configurations overcomes the energetic cost, which remains modest due to the soft  interparticle interaction~\cite{jacquinAnomalousStructuralEvolution2010}. 

As a final remark on the SAXS measurements, we note that for all samples the Kratky plots tend to flatten at $q > q_{max}$, $I(q)q^2$ being nearly constant over almost one decade in $q$. This points to a scale-free regime where scattering is dominated by ideal chain density fluctuations\cite{fischerWhenDoesMacromolecule2022,upadhyaPETRAFTSAXSHigh2019}. In this regime, the Kratky plateau increase with increasing temperature and mass concentration, compare the data at various $c$ of Figs.~\ref{fig:struct}a-c and the curves at various $T$ within each panel. This trend reflects an increase of the local monomer density, and a consequent decrease of the range of the spatial correlations of density fluctuations as the microgel VPT is approached.

\begin{figure}[htbp!]
\includegraphics[width=10cm]{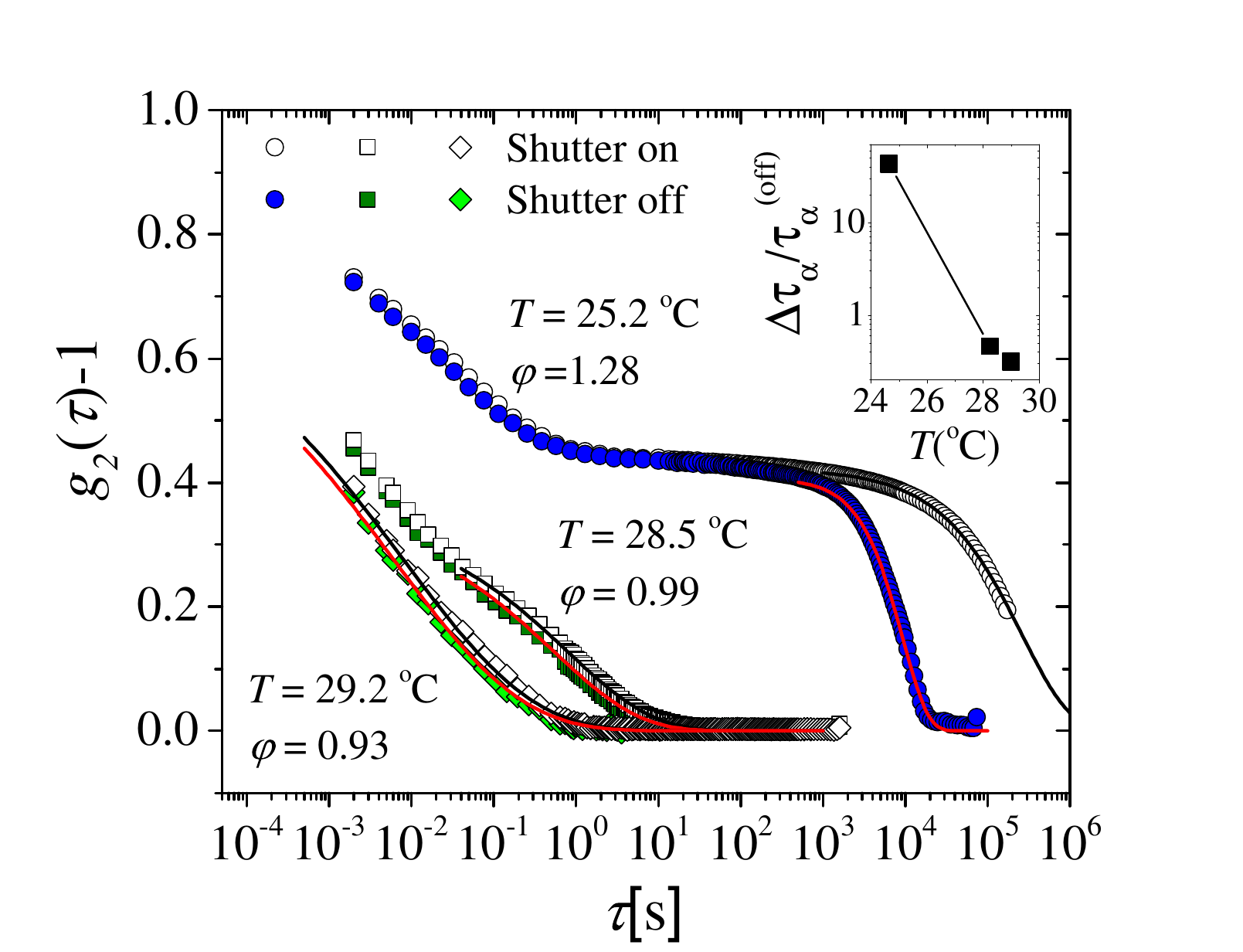}
\caption{\label{fig:shutter} Effect of reducing the average illumination power on the dynamics of the ULC2 microgels, for various $T$.  For the measurements with a shutter, the incident beam is blocked $92\%$ of the time, corresponding to an effective reduction of the power on the sample of a factor of 12.5. Inset: relative change of the relaxation time measured with and without the shutter, as a function of temperature.}
\end{figure}

The decrease of the intensity peak reported in in Fig.~\ref{fig:struct2}b is strongly reminiscent of the structural ``glass melting'' observed is simulations and theory for soft colloids~\cite{berthierIncreasingDensityMelts2010,delmonteNumericalStudyNeutral2024}; however, it is not accompanied by a speed-up of the dynamics by several orders of magnitude, as in Ref.~\cite{berthierIncreasingDensityMelts2010}. This  leaves open the question of the origin of the anomalous dynamical regime at $\phi \geq \phi_g$, where $\tau_{\alpha}$  nearly plateaus. Thermal tests on the COLIS setup serendipitously hinted to a (partial) explanation of these dynamics. It was noted that for PNIPAM-based samples the power absorbed by the $T$ control of the setup slightly changed upon switching on or off the optical laser, unlike for tests where the sample cell was filled with pure water. This suggests that PNIPAM may slightly absorb laser light, a quite surprising observation, since the primary UV-Vis absorption maximum for PNIPAM is around 228 nm \cite{makharzaStructuralThermalAnalysis2010}, far from the 532 nm wavelength of the optical laser. Light absorption would locally heat the suspension, potentially  accelerating the dynamics. 

To check whether illumination with visible light impacts the ULC dynamics, we repeated measurements at several $\phi$'s, using two distinct protocols, either illuminating the sample continuously, or using a shutter to block the incident beam except during image acquisition, thereby reducing the average sample illumination by about 92\%.   
The relaxation times obtained using continuous and reduced illumination are shown in Fig.~\ref{fig:tau vs phi}a,b as blue triangles and squares, respectively. We find that the dynamics are essentially insensitive to the illumination protocol in the supercooled regime. By contrast, for $\phi \gtrsim \phi_g$ continuous illumination results in a dramatic speed up of the dynamics, as compared to data collected using the shutter. Figure~\ref{fig:shutter} illustrates this effect by showing three pairs of intensity autocorrelation functions measured on exactly the same samples, using  the two protocols. The $g_2-1$ functions measured with either protocol are essentially identical for the two samples at lower $\phi$ (relative volume fraction $\phi/\phi_g = 0.823$ and $0.876$, respectively). Accordingly, the corresponding relative variation of the relaxation time, $\Delta \tau_{\alpha}/\tau_{\alpha}^{\mathrm{(off)}}$, is significantly smaller than unity, see the inset. [Here, $\Delta \tau_{\alpha} = \tau_{\alpha}^{\mathrm{(on)}} - \tau_{\alpha}^{\mathrm{(off)}}$, and the superscript indicates whether the shutter was used or not]. On the contrary, for the glassy sample at $\varphi=1.28$ not only do the dynamics slow down by a factor of more than 40 when using the shutter (inset of Fig.~\ref{fig:shutter}), the shape of $g_2-1$ also changes, since $\beta = 1.35$ and $0.65$, without and with the shutter respectively.

The experiments with the shutter suggest that the relaxation processes in the $\tau_{\alpha}$ plateau regime observed at $\phi > \phi_g$ are distinct from those at play in the supercooled regime. For $\phi < \phi_g$, the usual structural relaxation processes, e.g. the cage escape process described by the mode coupling theory~\cite{gotze_relaxation_1992}, are the fastest ones and set the time scale of $\tau_{\alpha}$. For very dense suspensions, above $\phi_g$, we speculate that structural relaxation via thermal motion becomes so slow that any additional relaxation mechanism is likely to prevail, ultimately setting the time scale of $\tau_{\alpha}$ . For our ULC microgels, absorption of laser light introduces additional dynamics via two contributions: on the one hand, a local increase of $T$ decreases $\phi$ in the scattering volume, due to particle shrinkage, hence speeding up the dynamics. On the other hand, it induces $T$- and $\phi$-gradients that may trigger mesoscopic flows of particles, which would further enhance dynamics.

Two final comments on the dynamics at very high $\phi$ are in order. First, it is unlikely that beam absorption alone is responsible for the anomalous dynamics often observed in very concentrated suspensions of soft colloids: indeed, similar dynamics have been reported for a variety of systems~\cite{li_long-term_2017,philippe2018glass,nigroRelaxationDynamicsSoftness2020,frenzel_glassliquid_2021}, with different composition, and illuminated with radiation with various $\lambda$, including for silica spheres that do not absorb significantly laser light~\cite{philippe2018glass}. Additionally, we note that for the experiment reported here reducing drastically the average illumination did not remove the slow relaxation, it just pushed it to longer time scales. Second, the scenario proposed here implies that the dynamics observed for our very concentrated ULC suspensions are different in nature than those reported in the framework of the reentrant glass transition of soft particles discussed in simulation and theoretical papers~\cite{berthierIncreasingDensityMelts2010,jacquinAnomalousStructuralEvolution2010,delmonteNumericalStudyNeutral2024}. One would thus be left with similarities in the structural features (the increase of disorder discussed in relation to Figs.~\ref{fig:struct} and~\ref{fig:struct2}), but different dynamics, calling for more experiments and theoretical developments to reach a deeper understanding of these systems.

\begin{figure}[htbp!]
\includegraphics[width=9cm]{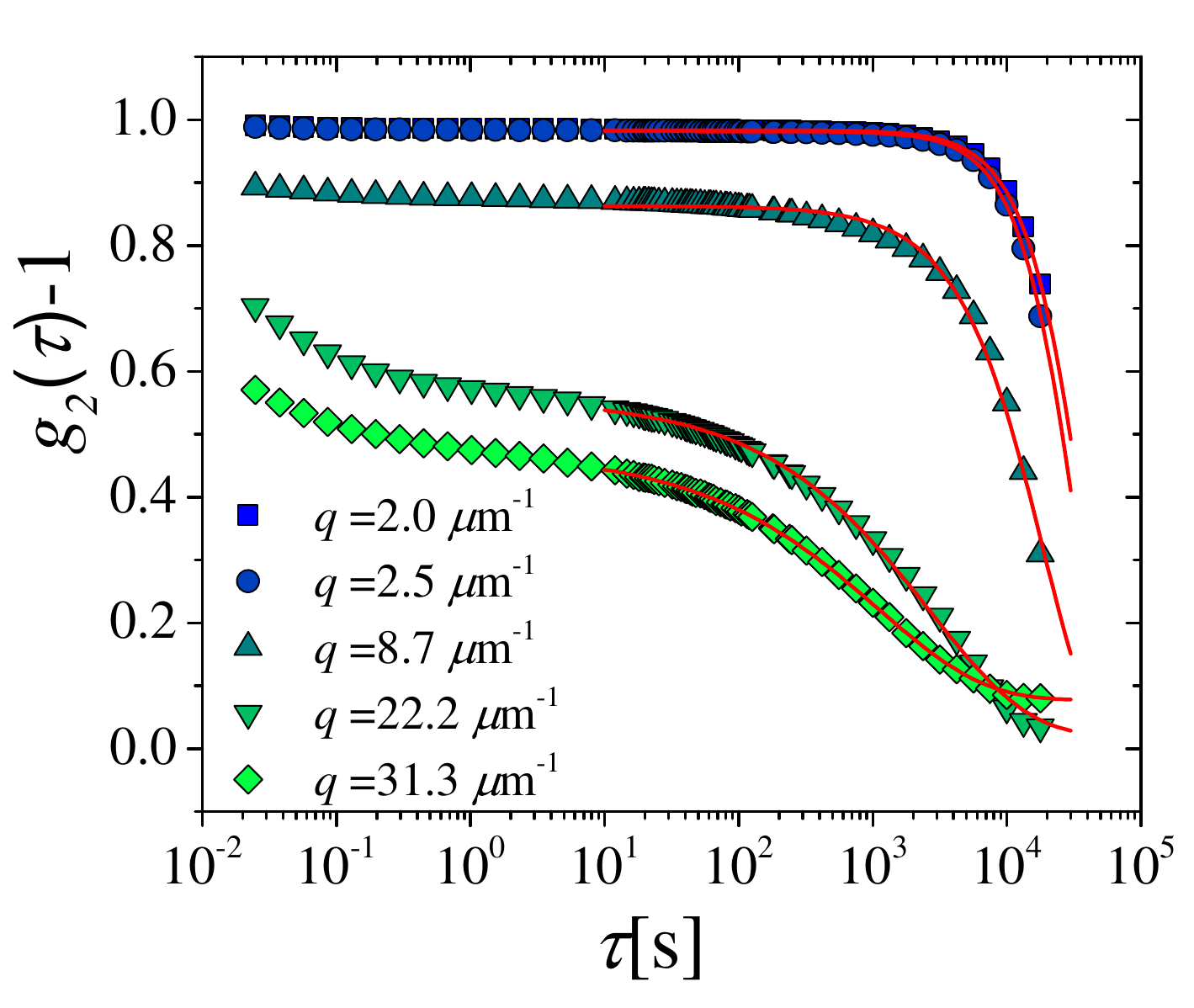}
\caption{\label{fig:qdep} Intensity correlation functions for the ULC2 microgels at $T=27.0~^{\circ}$C ($\varphi=1.13$), for various scattering vectors as shown by the labels,  measured with the COLIS setup, with the shutter. The red lines are KWW fits, Eq.~\ref{eq:KWW}, to the final relaxation of $g_2-1$. 
}
\end{figure}

\begin{figure}[hbp!]
\includegraphics[width=9cm]{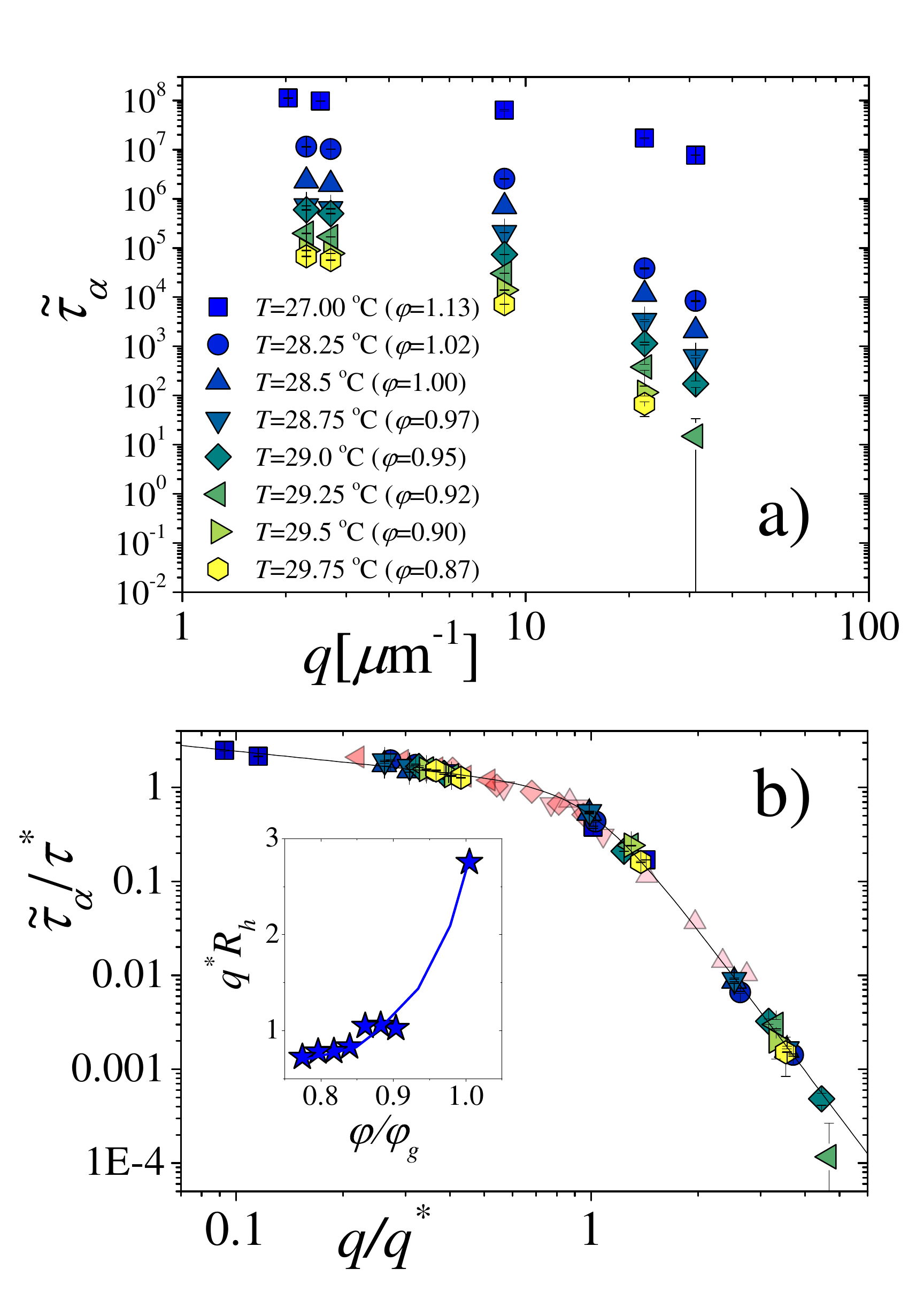}
\caption{\label{fig:qdep2} a: $q$-dependence of $\widetilde{\tau}_{\alpha}$, the relaxation time normalized by its $\phi \rightarrow 0$ value at $q = 22.2~\mu\mathrm{m}^{-1}$, to factor out the $T$ dependence of the microgel hydrodynamic radius and solvent viscosity. Data for the ULC2 sample probed by the COLIS setup (shutter active), at $c$=0.0335 wt/wt and various $T$, as indicated by the labels. b) Master curve obtained by plotting the data in a) (same symbols), using scaled variables, see text for details. Symbols with red-pink light shades: data from the microgels of Ref.~\cite{nigroRelaxationDynamicsSoftness2020} (
up triangles: $c_w$ = 0.004; down triangle:s $c_w$ = 0.008; diamonds: $c_w$ = 0.028; left triangles: $c_w$ = 0.04). The line is a fit to the data with Eq.~\ref{eq:master_curve}. Inset: $\phi$-dependence of the scaling scattering vector $q^*$. The line is a guide to the eyes.
}
\end{figure}

Further insight on the origin and nature of the dynamics may be gained by investigating their length-scale (or $q$) dependence, a feature explored in very few works so far, in particular a different microgel system~\cite{nigroRelaxationDynamicsSoftness2020}, and silica hard colloids coated with a PNIPAM layer~\cite{frenzel_glassliquid_2021}. Figure~\ref{fig:qdep} shows representative $g_2-1$ functions measured simultaneously at four $q$ vectors (COLIS setup, ULC2 sample, shutter activated), at $\phi = 1.13 =\phi_g$. At all $q$, the final decay is well fitted by the KWW function already introduced to fit the data at  $q=22.1~\mu \mathrm{m}^{-1}$, Eq.~\ref{eq:KWW}, although the fits somehow deviate from the data at the smallest $\tau$ of the fitting range. We find that $\beta$ decreases systematically with $q$, from $\beta \simeq 1.7$ at the smallest $q$, corresponding to a compressed exponential relaxation, to $\beta \simeq 0.6$ at the largest $q$, indicative of a stretched exponential relaxation.

Figure~\ref{fig:qdep2}a shows the $q$ dependence of the relaxation time extracted from the KWW fits for the same sample, at various $T$ corresponding to $0.87 \leq \phi \leq 1.13$, in the supercooled regime and up to $\phi_g$, beyond the regime where the VFT fit of $\tau_{\alpha}$ introduced in Fig.~\ref{fig:tau vs phi}a applies. Note that both the solvent viscosity and the microgel radius vary with $T$: we factor out this trivial dependence by plotting $\tilde{\tau}_{\alpha}$, the relaxation time normalized by its $\phi \rightarrow 0$ limit at $q=22.1~\mu \mathrm{m}^{-1}$. The corresponding values of $\beta$ are reported in the \SI. Two regimes are clearly seen: a weak $q$ dependence at low $q$, and a steeper decrease with $q$ at larger scattering vectors. 
Note that the two distinct regimes seen for our microgels are not compatible with the single power law $\tilde{\tau}_{\alpha} \sim q^{-p}$ behavior reported in Refs.~\cite{nigroRelaxationDynamicsSoftness2020,frenzel_glassliquid_2021}. 

A clear trend emerges from Fig.~\ref{fig:qdep2}a: upon increasing $\phi$, both the low- and high-$q$ regimes exhibit a milder $q$ dependence. This suggests that the data may be rescaled onto a master curve, by normalizing both $\tilde{\tau}_{\alpha}$ and $q$ by suitable $\phi$-dependent factors $\tau^*$ and $q^*$, respectively. The resulting master curve is shown in Fig.~\ref{fig:qdep2}b: an excellent collapse is seen not only for the data of the present work (same symbols as in Fig.~\ref{fig:qdep2}a), but also for the microgels of Ref.~\cite{nigroRelaxationDynamicsSoftness2020} (symbols with red and pink light shades, see caption). This suggests a scenario different from that proposed by Nigro et al., who hypothesized two different regimes, below and above the glass transition, respectively~\cite{nigroRelaxationDynamicsSoftness2020}. The master curve of Fig.~\ref{fig:qdep2}b rather points to a universal behavior throughout the supercooled and glass regime, with the experimentally accessible $q$ window shifting towards smaller (non-dimensional) scattering vectors $q/q^*$ as $\phi$ increases. Indeed, we find that $q^*$ increases with $\phi$, see the inset of Fig.~\ref{fig:qdep2}b. The growth of $q^*$ is modest in the supercooled regime, while a significantly larger value is observed for the most concentrated sample. The scaling parameter $\tau^*$ essentially accounts for the strong increase of the relaxation time with $\phi$ in dense suspensions: we find that $\tau^*(\phi)$ closely follows the growth of $\tau_{\alpha}(\phi)$, although with some deviations at the highest tested $\phi$, see Fig. S6 in the~\SI. Finally, we note that the relevance of the master curve for the data of Ref.~\cite{nigroRelaxationDynamicsSoftness2020} is confirmed by the fact that $q^*$ follows a trend similar to that of the inset of Fig.~\ref{fig:qdep2}b, see the \SI.

The master curve of Fig.~\ref{fig:qdep2}b may be fitted by various simple empirical functions accounting for the crossover between the weak $q$ dependence for $q<q^*$ and the steeper decay at larger $q$, including an exponential decay $\tilde{\tau}_{\alpha} \sim \exp[-(q/q^*)/\xi ]$ or a generalized Ornstein-Zernike function, $\tilde{\tau}_{\alpha} \sim [1+(\xi q/q^* )^2]^{-b}$, with no significant differences in the fit quality. A more physically-motivated choice should leverage the observation that the cross-over between the weak and strong $q$-dependent regimes occurs on a length scale $1/q^*$ comparable to or smaller than the microgel radius $R_h$, see the inset of Fig.~\ref{fig:qdep2}b. This suggests a hierarchy of distinct relaxation mechanisms involving the whole microgel or its internal structure, respectively. Inspired by this observation, we propose the following expression:  
\begin{equation}
\label{eq:master_curve}
    \widetilde{\tau}_{\alpha}/\tau^* = \frac{1}{\left(\frac{q}{q^*}\right)^a + \left(\frac{q}{q^*}\right)^b} \,,
\end{equation}
which fits very well the data with $a=0.39 \pm 0.02$ and $b=5.01 \pm 0.08$. Equation~\ref{eq:master_curve} has a particularly simple interpretation for the case of exponential relaxations, where the average relaxation rate, $1/\widetilde{\tau}$, is obtained from the weighted sum of two relaxation rates, $\Gamma_l \sim q^a$ and $\Gamma_h \sim q^b$, dominating at low and high $q$, respectively. In the more general case of non-exponential relaxations, which applies to our system, Eq.~\ref{eq:master_curve} still accounts for the overall relaxation time in terms of the contributions of two relaxation processes. 

Equation~\ref{eq:master_curve} suggests that the final decay of the intensity correlation functions should be fitted by an expression accounting for two processes. Two possible choices are the squared sum of KWW functions, commonly used for glassy dynamics~\cite{sidebottom1993two,de2015structural,nigroRelaxationDynamicsSoftness2020}, or the product of two such functions~\cite{frey2025liquid}, the latter choice corresponding to statistically independent processes with additive mean squared displacements~\cite{aime_microscopic_2018,busch2008dynamics}. Both functional forms work well; we choose the latter, which allows for more robust fitting. We thus fit the final decay of $g_2-1$ to the expression
\begin{equation}
\label{eq:productKWW}
    g_2 (\tau)-1\propto \exp\left [-2(\tau/\tau_{l})^{\beta_l}\right]\exp\left [-2(\tau/\tau_{h})^{\beta_h}\right] + B_{ln} \,, 
\end{equation}
where the subscripts $l$ and $h$ stand for the low- and high-$q$ regimes, respectively. Equation~\ref{eq:productKWW} fits very well the data, including at the onset of the final decay of $g_2-1$, where the single KWW function deviates from the data (compare Fig. SM11 in the~\SI~to Fig.~\ref{fig:qdep}).

\begin{figure}[htbp!]
\includegraphics[width=9cm]{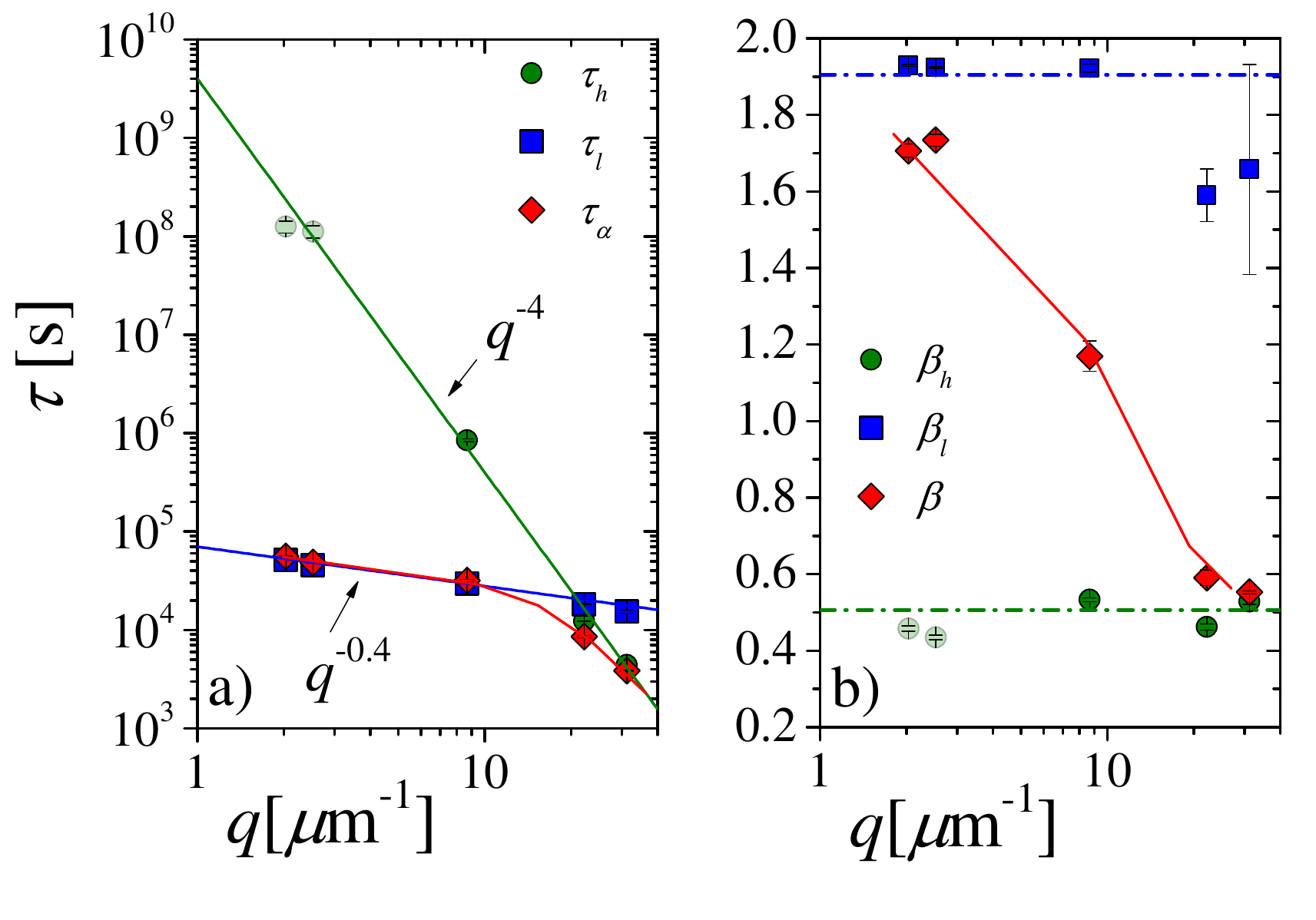}
\caption{\label{fig:fitparams_doubleKWW} Fitting parameters for the ULC2 sample at $\phi = 1.13$, obtained using a single KWW function, Eq.~\ref{eq:KWW} and red diamonds, or the product of two KWW functions, Eq.~\ref{eq:productKWW} and blue squares and green circles for the modes dominating at low and high $q$'s, respectively. a): relaxation time; b): shape parameter. In b, the dashed-dotted line are the $\beta$ values obtained by imposing a single value for each mode, shared among data at all $q$ vectors.}
\end{figure}

We show in Fig.~\ref{fig:fitparams_doubleKWW} the $q$ dependence of the fitting parameters obtained by fitting the $g_2-1$ functions of Fig.~\ref{fig:qdep} with Eq.~\ref{eq:productKWW}. Consistent with the shape of the master curve of Fig.~\ref{fig:qdep2}b, we find that the two processes have very different $q$ dependencies: $\tau_l \sim q^{-0.4}$ and $\tau_h  \sim q^{-4}$, respectively. As a result, the fitted relaxation times may differ by several orders of magnitude, making the determination of the longest relaxation time difficult at the lowest $q$'s. Points where the fit parameters associated to the slowest mode should be taken with caution are indicated by semi-filled symbols. The two modes have also distinct shape parameter $\beta$, see the green circles and blue squares in Fig.~\ref{fig:fitparams_doubleKWW}b: the low-$q$ mode is compressed, $1.6 \lesssim \beta_l \lesssim 1.9$, while the high-$q$ mode is stretched, $\beta_h \approx 0.5$. Within each mode, the variation of $\beta$ with $q$ is small and not systematic, suggesting that a fit of almost equal quality may be obtained using a single value for $\beta_l$ and one for $\beta_h$, shared among all data for various $q$'s. These values are shown by the dash-dotted lines in Fig.~\ref{fig:fitparams_doubleKWW}b; in fact, both the fits displayed in Fig. SM11 in the~\SI~and the relaxation times of Fig.~\ref{fig:fitparams_doubleKWW}a were obtained using shared values of the shape parameters. For comparison, Fig.~\ref{fig:fitparams_doubleKWW} shows also the fitting parameters obtained, for the same set of data, with the single KWW function used in the rest of the paper (red diamonds). When fitting with a single KWW function, one observes a progressive increase of the steepness of $\tau_{\alpha}$ \textit{vs} $q$, as well as a continuous decrease of $\beta$ with $q$. This behavior was also reported for the microgels of Ref.~\cite{nigroRelaxationDynamicsSoftness2020}. In the framework of the master curve proposed here, however, this smooth variation is a crossover effect: at small $q$, the dynamics are dictated essentially only by the low-$q$ mode (blue symbols in Fig.~\ref{fig:fitparams_doubleKWW}), while the importance of the high-$q$ mode (green symbols) grows with $q$, until it prevails at the largest scattering vector. Finally, we stress that the behavior discussed here for the sample at $\phi = 1.13$ also applies for the other samples of Fig.~\ref{fig:qdep2} at lower $\phi$, although in practice for several $q$'s one single mode dominates the relaxation of $g_2-1$. The $q$ dependence of the exponents $\beta$ for the two processes and for all samples is shown in the\SI. 

The scenario suggested by Figs.~\ref{fig:qdep2} and~\ref{fig:fitparams_doubleKWW} is the following: there are two distinct dynamical processes, dominating at low and high $q$'s, respectively, with a crossover length scale comparable to the microgel size. Small $q$ dynamics are weakly $q$ dependent, and are characterized by a shape parameter $\beta \geq 1.6$. Both features have been reported for the low-$q$ dynamics of colloidal gels formed by attractive particles~\cite{cipellettiUniversalAgingFeatures2000,choEmergenceMultiscaleDynamics2020,bouzidElasticallyDrivenIntermittent2017}, where they have been ascribed to the slow relaxation of internal stress built-in upon quenching the system in states with extremely slow dynamics. Our data suggest that a similar scenario also applies to repulsive colloidal glasses. On length scales smaller than the microgel size, by contrast, the relaxation time decreases steeply with $q$: $\tau_h \sim q^{-5}$, indicative of strongly sub-diffusive dynamics. As also suggested by Nigro \textit{et al.}~\cite{nigroRelaxationDynamicsSoftness2020}, it is likely that this behavior is related to the dynamics of the polymer chains that form the microgel. Indeed, we observe that while the master curve introduced here rationalizes both our data and those of the microgels of Ref.~\cite{nigroRelaxationDynamicsSoftness2020}, it is not applicable to the PNIPAM-coated silica particles of Ref.~\cite{frenzel_glassliquid_2021}, where the polymer layer contributes very weakly to the XPCS scattering signal, which is dominated by the hard silica core. 

We note that the $\tau_h \sim q^{-5}$ scaling inferred from the master curve is not compatible with the predictions of the Rouse ($\tau\sim q^{-2}$) or Zimm ($\tau\sim q^{-3}$) models \cite{hsuDetailedAnalysisRouse2017,doiTheoryPolymerDynamics1986} for coherent scattering. However, for glassy polymers, both numerical simulations~\cite{roe_molecular_1994,c._benneman_growing_1999} and quasi-elastic neutron scattering experiments~\cite{colmenero_correlation_1992} have reported strong $q-$dependent dynamics in the small $q$ range. In these works, the $\alpha$ relaxation time was found to scale as $q^{-b}$, with $b$ in the range 3.8 -- 9.5 depending on the system. Furthermore, the shape parameter $\beta$ in Ref.~\cite{colmenero_correlation_1992} varied between 0.21 and 0.44, a stretched exponential behavior reminiscent of $\beta_h \approx 0.5$ as found for our microgels. We note however that there are important differences between these works, which deal with glassy melts of polymers, and the cross-linked, solvent-swollen internal structure of the microgels: more work will be needed to fully understand the origin of the high $q$ regime in concentrated microgel suspensions. Finally, we recall that the crossover $q^*$ between the the small-$q$, stress-relaxation driven regime and the high-$q$ one shifts towards smaller length scales upon increasing $\phi$ (inset of Fig.~\ref{fig:qdep2}b). In the scenario described above, this would result from the polymer chains within the microgels being increasingly constrained by the sample compression, such that the polymeric degrees of freedom may only emerge on increasingly smaller length scales.   

\section{Conclusions}\label{sec:CC}
We have investigated the structure and dynamics of dense suspensions of ULC microgels with small-angle X ray scattering and PCI, a dynamic light scattering technique. The control parameter is the effective volume fraction $\phi$, which we varied both by changing the number density $n$ of microgels and the microgels' volume, leveraging the $T$-controlled swelling and deswelling of these colloids. In the highly supercooled and mildly glassy regimes, the structure depends separately on $n$ and $T$. By contrast, the dynamics only depend on $\phi$: since a priori one may expect that the microgels' softness --and hence their interaction potential-- depends on the swelling state and on the osmotic pressure exerted by the suspension itself and the associated counterions, this result is quite remarkable. It extends to the microscopic dynamics of ULC microgels previous findings for the macroscopic viscosity of conventional microgels~\cite{sessomsMultipleDynamicRegimes2009}, thus indicating that the effective volume fraction is a robust parameter dictating the flow and dynamical properties of microgel suspensions. It should be noted that usually the structure and the dynamics of dense suspensions are closely related to each other, a relationship upon which theories such as the mode coupling theory are built~\cite{gotze_relaxation_1992}. Furthermore, small changes in the structure typically result in order-of-magnitude variations of the relaxation time. It is therefore quite surprising that samples with very different microstruture, obtained by varying $n$ or $T$, but with the same $\phi$ exhibit the same dynamics. A deep understanding of this result will require further work, including numerical. 

In the deep supercooled and mildly glassy regimes, the decrease of local order with increasing $\phi$ revealed by SAXS is reminiscent of the re-entrant melting scenario unveiled in simulations of model soft particles~\cite{berthierIncreasingDensityMelts2010} and realistic conventional microgels~\cite{delmonteNumericalStudyNeutral2024}(note however that in very recent simulations of ULC microgels no structural re-entrance was seen~\cite{marin-aguilar_unexpected_2025}). In simulations, this melting is accompanied by an acceleration of the dynamics~\cite{berthierIncreasingDensityMelts2010} or, on the contrary, by a steady slowing down with increasing packing~\cite{delmonteNumericalStudyNeutral2024}. Neither behavior is observed for our microgels. Instead, at high $\phi$ the ULC microgels exhibit a nearly $\phi$-independent relaxation time, similarly to conventional microgels~\cite{philippe2018glass} and reminiscent of the flattening of $\tau_{\alpha}(\phi)$ for polymer-coated particles~\cite{li_long-term_2017,frenzel_glassliquid_2021}. We have shown that the slight absorption of laser light by PNIPAM is partially responsible for this effect. Light absorption causes a small local heating of the suspension, thus decreasing $\phi$ in the scattering volume and creating $\phi$ gradients across the sample, which are likely to contribute to the faster-than-expected dynamics observed in the glassy regime. More work will be needed to investigate systematically this effect, including its dependence on the radiation source, e.g. in XPCS experiments. A related question is that of the generality of this effect: we note that a similar plateau or flattening of $\tau_{\alpha}(\phi)$ at high $\phi$ has been reported for other soft repulsive colloids~\cite{li_long-term_2017,philippe2018glass,frenzel_glassliquid_2021}, including in suspensions of charge-stabilized silica particles. These colloids are not thermosensitive; furthermore the absorption coefficient of silica in the visible light range is lower than that of PNIPAM. Accordingly, local heating, if any, should have a negligible impact on the dynamics, leaving the question of the origin of these anomalous dynamics open.

At lower $\phi$, in the supercooled regime, we find that the increase of the relaxation time with $\phi$ is well described by a VFT-like law, implying that ULC microgels are fragile glass formers, similarly to hard spheres. Since the bulk modulus of ULC microgels is 10 times lower than that of conventional PNIPAM microgels, this is a strong indication that softness does not change change qualitatively the steep increase of $\tau_{\alpha}$ on approaching the glass transition. We note however that a very recent work on the same kind of ULC microgels reports a much gentler increase of the relaxation time and of the suspension viscosity with $\phi$~\cite{burger_suspensions_2025}.  While the $\tau_{\alpha}$ data of Ref.~\cite{burger_suspensions_2025} probably cover too small a range in relaxation times to allow for a definitive conclusion to be drawn, it is clear that the viscosity data are incompatible with a VFT-like increase of $\eta(\phi)$. This discrepancy may originate from slight differences in the synthesis protcol, leading to a larger mesh size in the microgels of Ref.~\cite{burger_suspensions_2025}: more work will be needed to fully elucidate this point.
%

Finally, we have found that in the supercooled and mildly glassy regime the dynamics measured at different $q$ vectors and $\phi$ obey a ``time-length scale superposition principle'' reminiscent of the time-temperature superposition principle in the mechanical or dielectric response of molecular glass formers~\cite{larson_structure_1998,olsen_time-temperature_2001}: all $\tau_{\alpha}(q,\phi)$ data fall onto a non-trivial master curve when using suitable reduced variables. The low-$q^*$  portion of the curve exhibits a very mild $q$ dependence of the dynamics; it describes data at very small $q$ vectors and/or high $\phi$. The reverse apply to the high-$q^*$ portion of the master curve, where an anomalously steep decrease of $\tau^*_{\alpha}(q^*)$ is seen. We proposed to rationalize this master curve by assuming that two distinct, hierarchical dynamical processes coexist, one implying the relaxation of density fluctuations via the collective motion of microgels, and the second one involving localized motion of the polymer chains within the microgels. This description, however, is not the only one possible: alternative scenarios could be based on a single dynamical process whose features smoothly evolve upon increasing $\phi$. More insight will probably come from numerical simulations, which are now able to include polymeric degrees of freedom~\cite{delmonteNumericalStudyNeutral2024,marin-aguilar_unexpected_2025}, and from the comparison with experiments on other soft particles. In this view, we find particularly interesting the finding that data for a different microgel system~\cite{nigroRelaxationDynamicsSoftness2020} can be scaled onto the same master curve, in contrast to measurements for non-polymeric particles~\cite{frenzel_glassliquid_2021}. Overall, we believe that the open questions briefly discussed above demonstrate how lively the research on microgels is and will probably continue to be in the next years.

\begin{acknowledgments}
We thank B. Guiselin, W. Kob, B. Ruta, V. Nigro, B. Ruzicka, and R. Angelini for illuminating discussions, G. Pr\'evot for help in instrumentation, and N. Segers and E. Leussink for help with the COLIS measurements. The SAXS measurements were performed on the CoSAXS beamline at the MAX IV laboratory (Lund, Sweden) under the proposals 20231840 and 20231866, the authors thank A. E. Terry and T. Plivelic for the support. AS acknowledges financial support from the Knut and Alice Wallenberg Foundation (Wallenberg Academy Fellows) and from the Swedish Research Council (Research Grant 2024-04178). RE and DT acknowledge financial support from the Agence Nationale de la Recherche (Grant ANR-20-CE06-0030-01; THELECTRA). AM was supported by the French CNES (Centre National d’Etudes Spatiales). LC gratefully acknowledges support by the French CNES, ESA and the Institut Universitaire de France. 
\end{acknowledgments}

\section*{Data Availability Statement}
All data used in the figures of the main text and the Supplemental Material are available on Zenodo [\textit{hyperlink to be inserted upon acceptance}].


\newpage
\onecolumngrid

\setcounter{figure}{0}
\makeatletter
\renewcommand{\thefigure}{S\@arabic\c@figure}
\makeatother

\section*{Supplementary Material}

\section*{Swelling curves}

\begin{figure}[htbp!]
    \centering
    \includegraphics[width=0.8\textwidth]{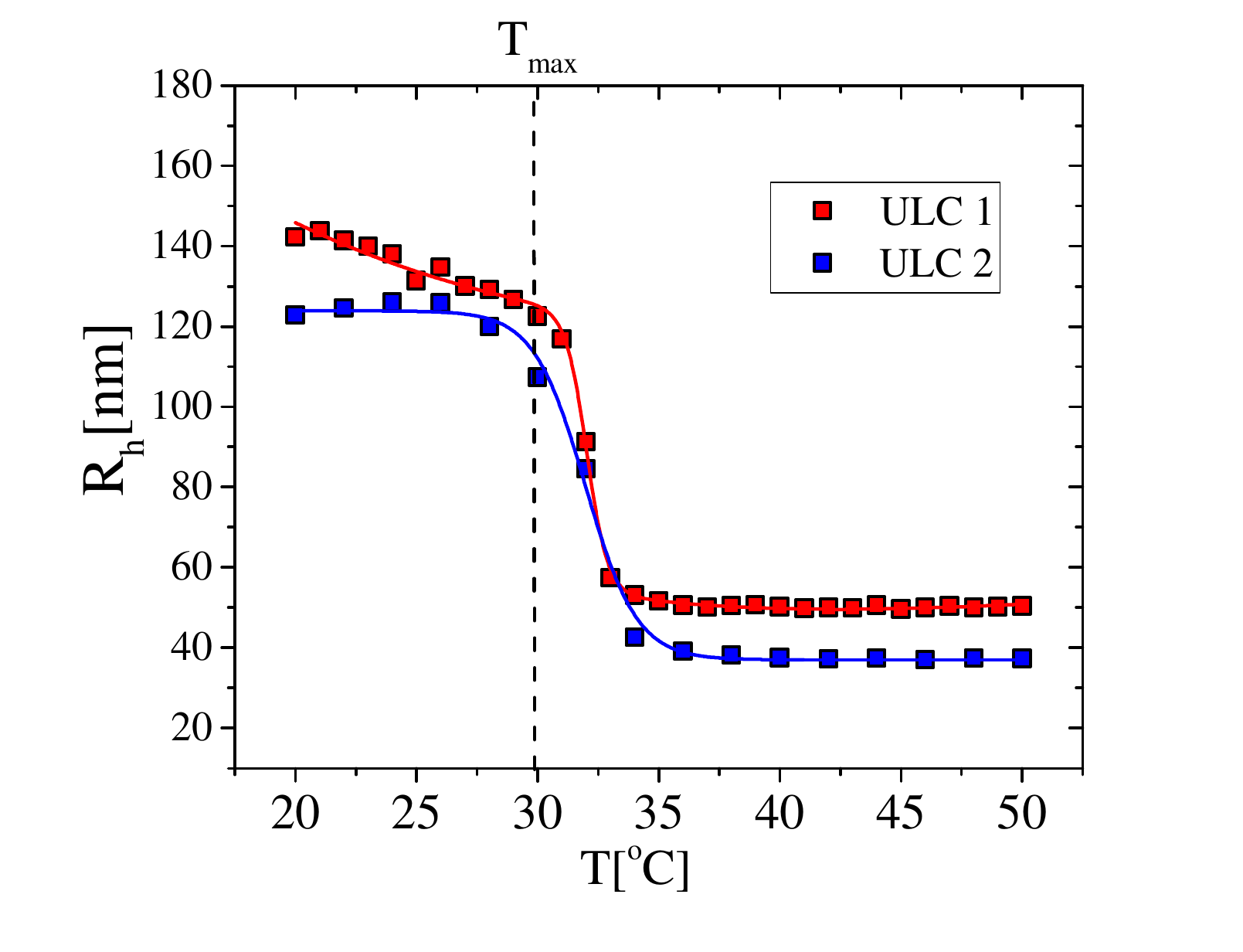}
    \caption{Swelling curves for the ULC1 and ULC2 microgels. The lines are the best fits of the experimental data $R_h(T)$ to Eq.~\ref{eq:Rh}. The vertical dashed line indicates to the maximum temperature reached in light scattering experiments on dense suspensions.}
    \label{fig:swelling}
\end{figure}

Figure~\ref{fig:swelling} shows the swelling curves for both ULC1 and ULC2 samples measured for very dilute samples ($c=0.0001$) by dynamic light scattering (DLS). The temperature dependence of the hydrodynamic radius $R_h$ has been fitted to obtain the radius at any temperature and obtain the microscopic time $\tau_0(T)=6\pi\eta(T)R_h(T)/(k_B T q^2)$, where $\eta(T)$ is the viscosity of water at temperature $T$, $k_B$ the Boltzmann constant, and $q=22.2~\mu$m$^{-1}$ the wave vector at which the swelling curves have been measured. The data have been fitted using the phenomenological function~\cite{delmonteTwostepDeswellingVolume2021a}

\begin{equation}\label{eq:Rh}
    R_h(T) = R_0-\Delta R \tanh[s(T-T_c)]+ A(T-T_c)+ B(T-T_c)^2 +C(T-T_c)^3, 
\end{equation} 
which is able to accurately capture the behavior of $R_h$ in the whole investigated temperature range.
Here $T_c$ is the volume phase transition temperature, $R_0$ is the radius of the microgel at the VPT, $\Delta R_h$ is the amplitude of the VPT, and the parameter $s$ quantifies its
sharpness.

\section*{Bulk modulus of ULC-microgels}

\begin{figure}[htbp!]
    \centering
    \includegraphics[width=0.8\textwidth]{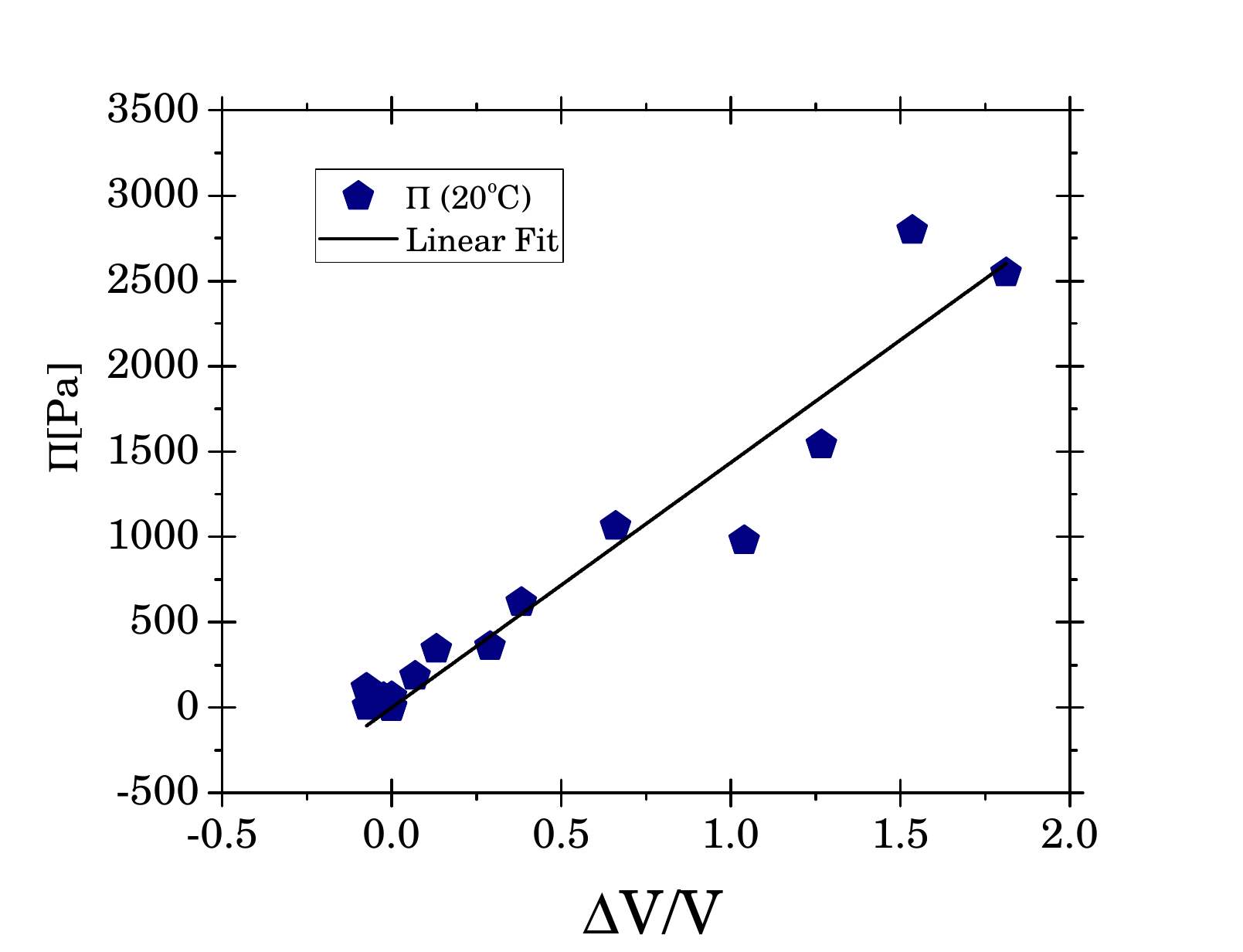}
    \caption{Osmotic pressure $\Pi$ as a function of the relative variation of the microgel volume, $\Delta V/V$, for ULC1 microgels}
    \label{fig:compression}
\end{figure}

In order to determine the bulk modulus of individual ULC1 microgels, we follow the method used in Ref.~\cite{philippe2018glass}, measuring by DLS the variation of the particle size upon applying an external osmotic pressure (Fig.~\ref{fig:compression}),
which is imposed by adding polyethylene glycol (PEG) with molecular weight Mw = 35 kDa.
Specifically, we measure the PNiPAM size using DLS for increasing concentrations
$c_{PEG}$ of the added polymer. To extract the microgel size from the DLS data, we use the viscosity
of the PEG35k solutions as measured by standard rheometry (for $c_{PEG}$ > 1:5 wt$\%$ ) or using
an Anton Paar Lovis 2000 ME microviscosimeter (at lower $c_{PEG}$). The osmotic pressure of PEG 35k solutions has been previously measured by Rami et al. \cite{ramiColloidalStabilityConcentrated2009} and Philippe et al. \cite{philippe2018glass}. It is fitted well in the range of interest by the equation proposed in Ref.~\cite{liEquationStatePEG2015}, though which we obtaine $\Pi$ for the desired $c_{PEG}$.
We calculate the compression modulus $K=-V\left(\frac{d\Pi}{dV}\right)$ of ULC1 microgels by performing a linear fit of the pressure, $\Pi=K\Delta V/V$, where $V=4/3\pi R_h^3$ is the microgel volume and $\Delta V$ is its variation due to osmotic compression. We obtain $K=1.43\pm 0.07$~kPa, in fair agreement with previous measurements of the ULC microgel modulus \cite{houstonResolvingDifferentBulk2022}, indicating that the bulk modulus of ULC microgels is about one order of magnitude lower than that of standard crosslinked PNIPAM microgels~\cite{philippe2018glass}.

\section*{Effective volume fraction}

In contrast to hard-sphere systems, microgels can adapt their size and shape due to changes in the suspension osmotic pressure and due to the contact with neighbouring particles.
To keep a common language with the case of hard spheres, where the sample concentration is quantified by the packing fraction $\varphi$, one can use the so-called generalized, or effective, volume fraction, $\zeta$, where the value of the particle volume $v_0$ is fixed to the volume of the microgels in the limit of highly diluted samples.  
The relation $\varphi = \zeta$ is true if the volume of the microgels do not change.
This condition is usually fulfilled at low concentrations. In the main text and in the following, we use $\varphi$ to indicate the effective volume fraction.

\begin{figure}[htbp!]
    \centering
    \includegraphics[width=0.8\textwidth]{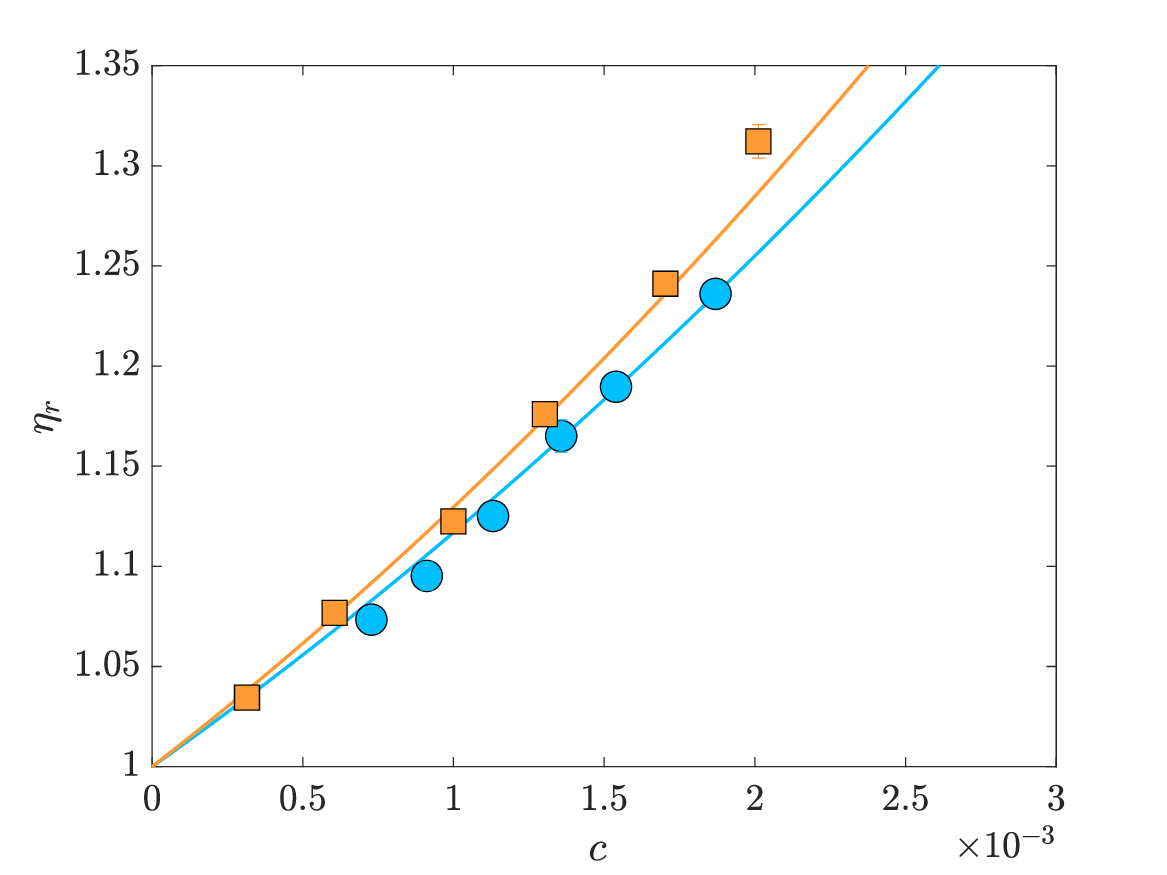}
    \caption{Relative viscosity $\eta_r$ \textit{vs} mass concentration $c$ (w/w) for diluted suspensions of ULC1 (circles) and ULC2 (squares) microgels. The solid lines are fits with Eq.~\ref{eq:EB2}.}
    \label{fig:visc}
\end{figure}

Increasing the packing fraction of a suspension of hard spheres results in an increase of the suspension viscosity. In the limit of highly diluted samples, $\varphi\ll 0.1$, the Einstein-Batchelor equation describes well the $\phi$ dependence of $\eta_r$, the suspension viscosity normalized by that of the solvent: 
\begin{equation}\label{eq:EB}
    \eta_r = 1 + 2.5\varphi +5.9\varphi^2\,.
\end{equation} 
Equation~\ref{eq:EB} holds also for suspensions of microgels in the limit of highly diluted samples. 
One can proceed with the substitution $\varphi = \zeta$, true in the limit of validity of the Einstein-Batchelor equation, and further note that $\zeta$ is proportional to $c$, the mass fraction of polymer in the sample (w/w), which is the experimental control parameter when preparing a suspension from a dried microgel powder. Therefore, Eq.~\ref{eq:EB} may be written as:
\begin{equation}\label{eq:EB2}
    \eta_r = 1 + 2.5kc +5.9(kc)^2\,, 
\end{equation} 
with $k$ a constant to be determined. Equation~\ref{eq:EB2} is used to fit the values of the relative viscosity of highly diluted solutions of microgels as a function of $c$ (sse Fig.~\ref{fig:visc}), yielding the desired conversion constant $k$. 
We used this procedure to determine $k$ for the two batches ULC1 and ULC2, as reported in the main text.

\section*{Equilibrium \textit{vs} non-equilibrium microscopic dynamics below and above $\varphi_g$}

\begin{figure}[htbp!]
    \centering
    \includegraphics[width=0.9\textwidth]{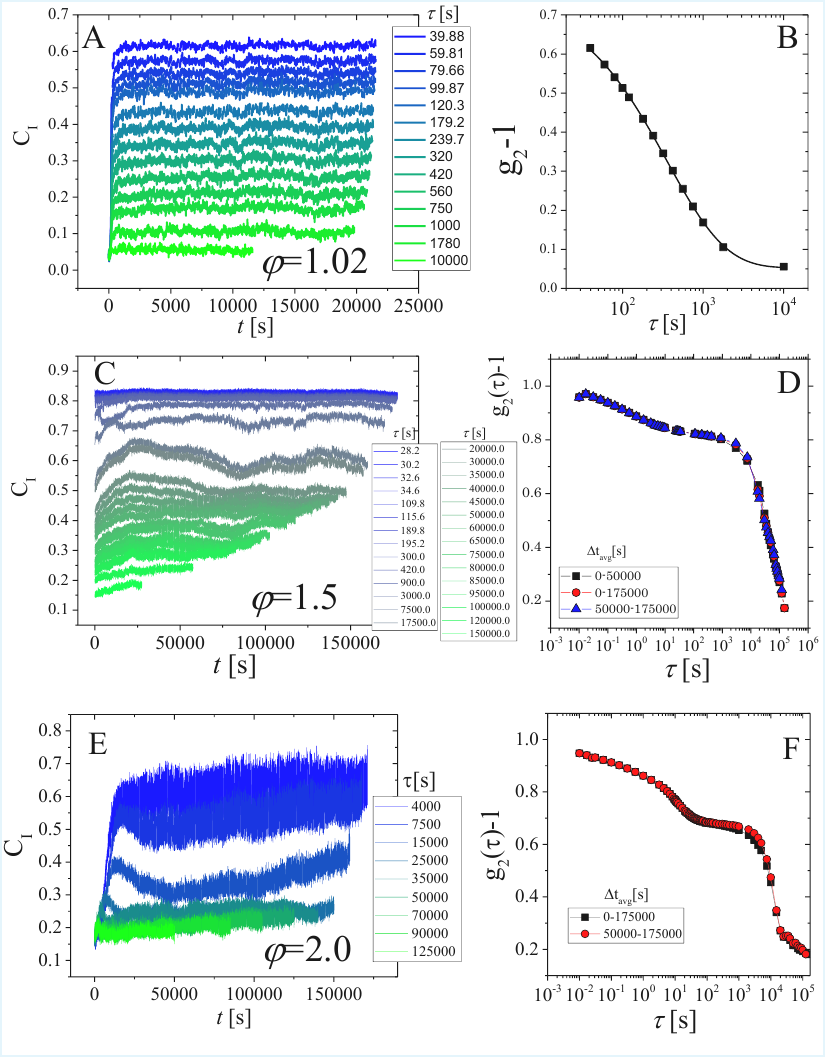}
    \caption{(A,C,E): two-time degree of correlation $C_I(\tau,t)$ defined in the main textas a function of time ($t$) and selected lag times $\tau$ as shown by the labels. (B,D,F); corresponding intensity autocorrelation functions $g_2(\tau)$-1 obtained by averaging $C_I(\tau,t)$ over the time intervals indicated by the labels. Data for three volume fractions of microgels as indicated in (A,C,E). The sample (ULC2) at $\varphi=1.02$ (Panels A,B) is below $\varphi_g^{(2)}=1.13$ and does not show ageing nor intermittent dynamics. The other two samples (ULC1) are well within the glassy regime ($\varphi > \varphi_g^{(1)}=1.00$), showing large fluctuations typical of incomplete equilibration and mild aging,  
    which has been also reported for standard crosslinked microgels above $\varphi_g$. \cite{philippe2018glass}}
    \label{fig:cI}
\end{figure}

Figure~\ref{fig:cI} shows the two-time degree of correlation $C_I$ defined in the main text and corresponding time-averaged correlation functions, both below and above the glass transition volume fraction, $\phi_g$. While samples below $\phi_g$ are fully equilibrated, at higher $\phi$ a mild aging is observed.

\clearpage
\section*{$q^*R_h$ for the data extracted from Nigro et al.\cite{nigroRelaxationDynamicsSoftness2020}}

\begin{figure}[htbp!]
    \centering
    \includegraphics[width=0.6\textwidth]{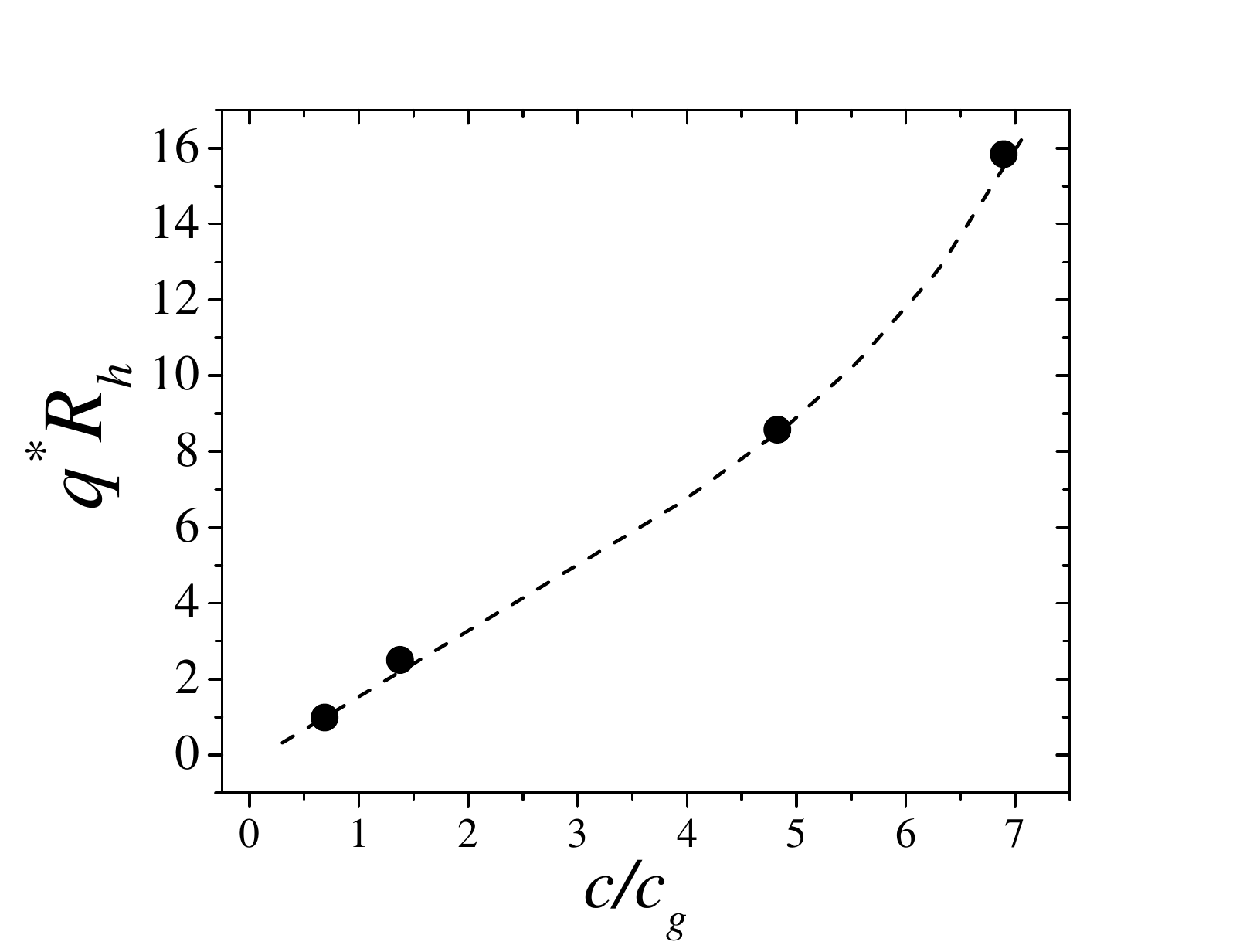}
    \caption{$\varphi$-dependence of the normalized scaling scattering vector $q^{*}R_h$ obtained from the data published in Ref.~\cite{nigroRelaxationDynamicsSoftness2020}. The line is a guide for the eye.}
    \label{qstar_Nigro}
\end{figure}

Figure~\ref{qstar_Nigro} shows the scaling $q$ vector used to collapse the data of Ref.~\cite{nigroRelaxationDynamicsSoftness2020} onto the master curve presented in Fig.~7b of the main text. The trend is similar to that observed for the ULC microgels of the present work (compare to the inset of Fig.~7b of the main text), although here the range of explored concentrations and the growth of $q^*$ are larger.

\clearpage
\section*{$\tau^*$-scaling}

\begin{figure}[htbp!]
    \centering
    \includegraphics[width=0.8\textwidth]{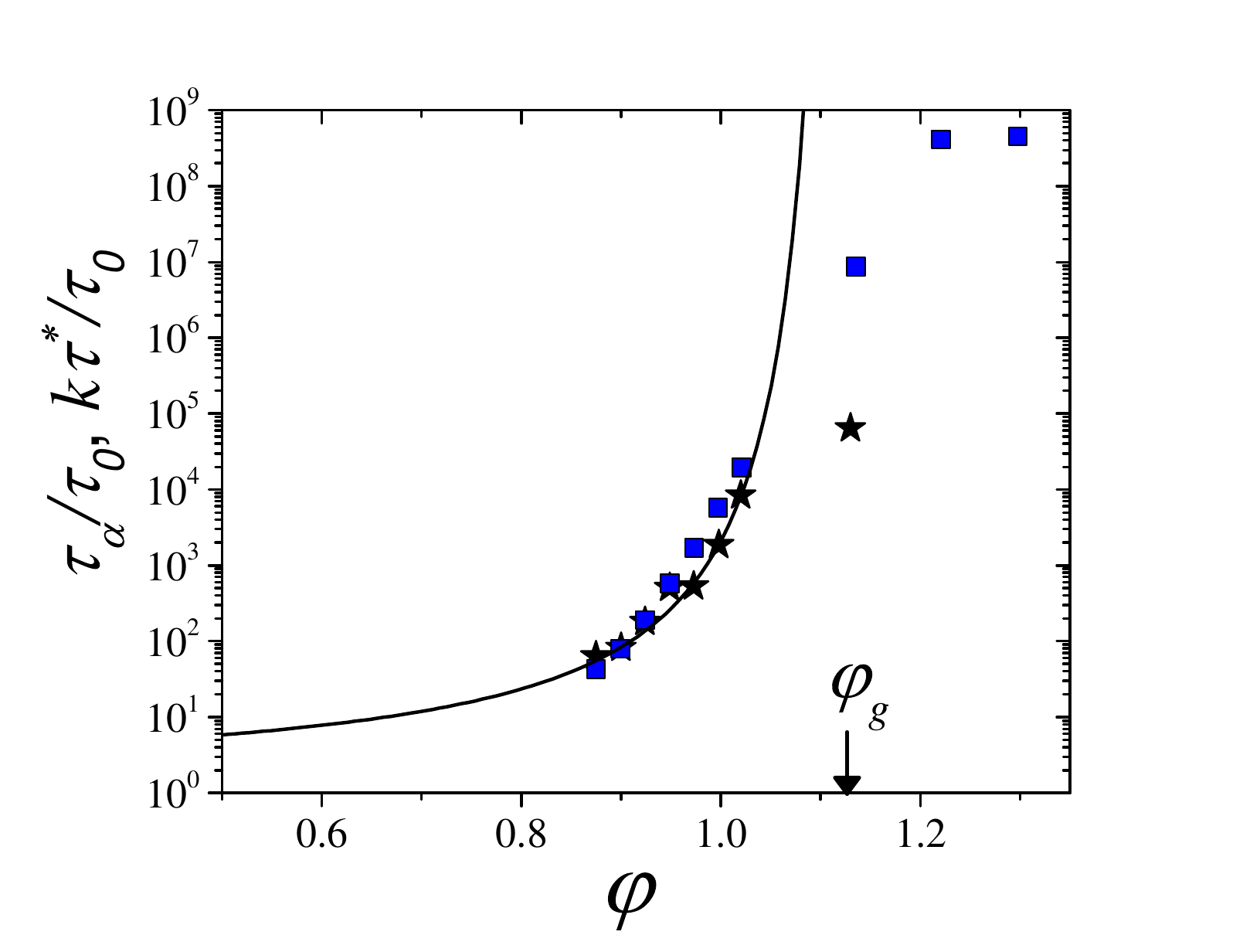}
    \caption{$\varphi$-dependence of the normalized scaling relaxation times $k\tau^{*}/\tau_0$ (black stars), with $k=1.43\cdot 10^{-3}$ a multiplicative constant to match the time units. The scaling times are plotted together with $\tau_{\alpha}/\tau_0$ (ULC2[$T$]) and the best VFT-fit of the data shown in Fig. 2 of the main text.}
    \label{fig:taustar_scaling}
\end{figure}

Figure~\ref{fig:taustar_scaling} compares the $\phi$ dependence of the time scale $\tau^*$ --used to collapse the data of the present work on the master curve shown in Fig.~7b of the main text-- to that of the $\alpha$ relaxation time measured at $\theta = 90~$deg for the ULC2 sample. Both follow the same trend up to $\phi \lesssim \phi_g$.

\clearpage
\section*{Master curve: comparison with data by Frenzel et al.~\cite{frenzel_glassliquid_2021} on dense suspensions of PNIPAM-coated silica particles.}

\begin{figure}[htbp!]
    \centering
    \includegraphics[width=0.8\textwidth]{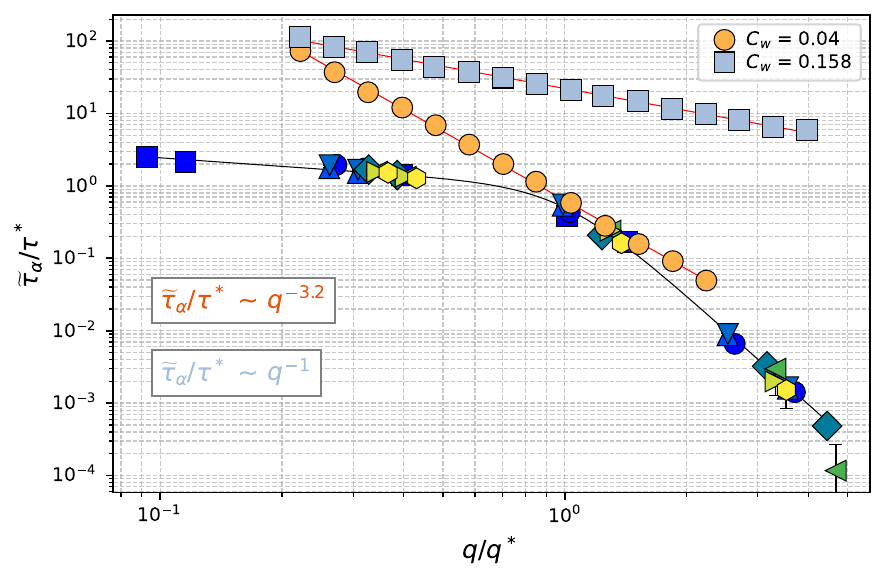}
    \caption{Master curve as reported in the main text together with the (rescaled) data of Frenzel et al. \cite{frenzel_glassliquid_2021} for two representative mass fraction concentrations $C_w$, in the repulsive glass regime of Ref.~\cite{frenzel_glassliquid_2021}. The red lines are best power-law fits to the data.}
    \label{fig:frenzel}
\end{figure}

Figure~\ref{fig:frenzel} shows that the data for concentrated suspensions of PNIPAM-coated silica particles~\cite{frenzel_glassliquid_2021} cannot be scaled onto the master curve established in Fig.~7b of the main text, unlike those for our ULC microgels and for the microgels of Ref.~\cite{nigroRelaxationDynamicsSoftness2020}.

\clearpage
\section*{Master curve fits}

\begin{figure}[htbp!]
    \centering
    \includegraphics[width=0.8\textwidth]{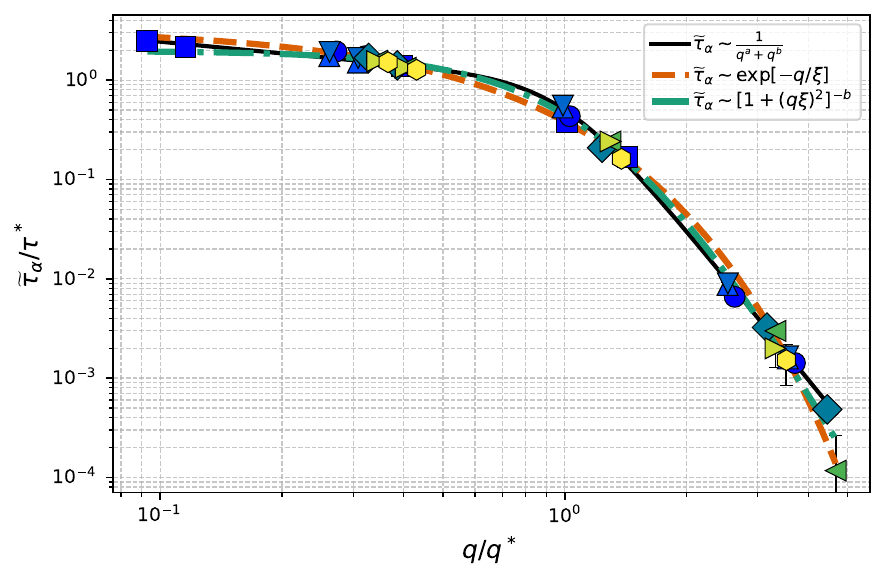}
    \caption{Empirical functions accounting for the crossover between the weak $q$ dependence for $q<q^*$ and the steeper decay at larger $q$: an exponential decay, $\tilde{\tau}_{\alpha} \sim \exp[-(q/q^*)/\xi ]$ (orange dashed line), a generalized Ornstein-Zernike function, $\tilde{\tau}_{\alpha} \sim [1+( \xi q/q^* )^2]^{-b}$ (green dashed line), and Eq.~6 of the main text (black continuous line).}
    \label{fig:master}
\end{figure}

Figure~\ref{fig:master} shows various fits to the master curve introduced in the main text (see therein for details).

\clearpage
\section*{q-dependence of the shape parameter $\beta$: single-KWW fits}

\begin{figure}[htbp!]
    \centering
    \includegraphics[width=0.8\textwidth]{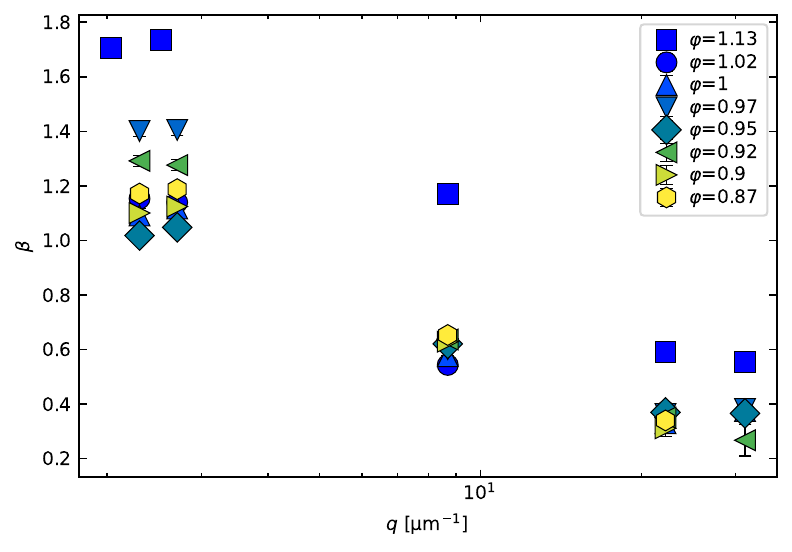}
    \caption{Shape parameter as a function of scattering vector for different volume fractions, as obtained from the single KWW fits.}
    \label{fig:beta_single}
\end{figure}

Figure~\ref{fig:beta_single} shows the $q$ dependence of the shape parameter $\beta$ issued by fitting the intensity correlation function using a single KWW decay, for various $\phi$ (ULC2 sample, shutter activated).

\clearpage
\section*{q-dependence of the shape parameter $\beta$: double-KWW fits}

\begin{figure}[htbp!]
    \centering
    \includegraphics[width=0.8\textwidth]{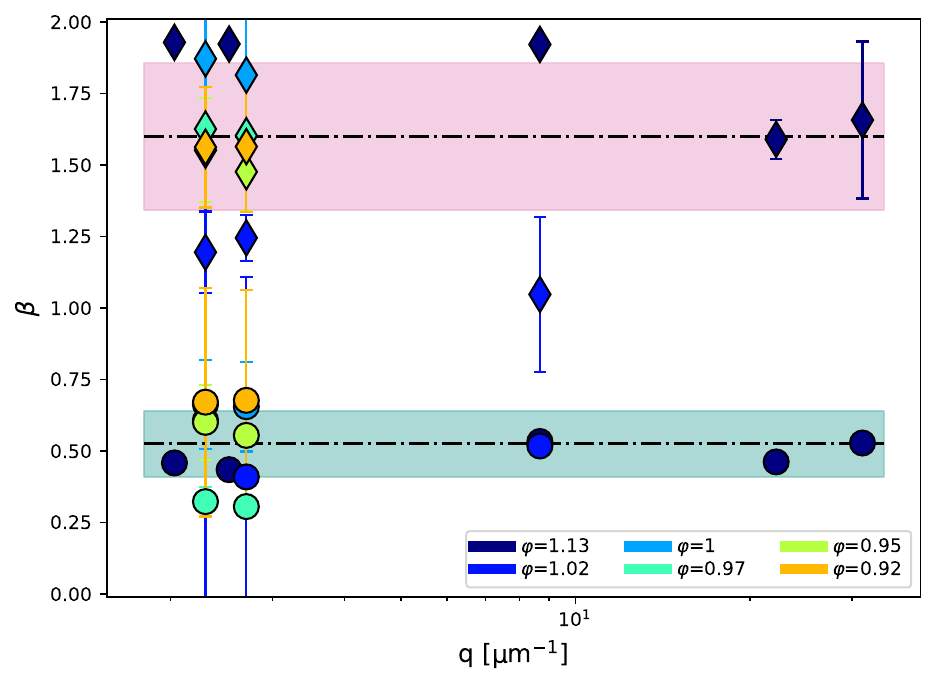}
    \caption{Shape parameter obtained fitting the data with Eq. 7 of the main text for different volume fractions, as shown in the legend. Diamonds (resp., circles) are the values for the compressed (resp., stretched) values, i.e. $\beta_l$ (resp., $\beta_h$) as discussed in the main text. For the low volume fractions ($\varphi\leq 1$), the fitting with a double exponential does not converge at high $q$-values, despite working well at low $q$: this is due to the fact that at high $q$ the compressed relaxation is too slow compared to the stretched one, and the curves display essentially only a single (stretched) decay. 
    }
    \label{fig:beta_double}
\end{figure}

Figure~\ref{fig:beta_double} shows the $q$ dependence of the shape parameter $\beta$ issued by fitting the intensity correlation functions using the product of two KWW decays, as discussed in the main text. The horizontal lines show the $\beta$ values obtained by imposing values of $\beta$ shared among data at all $q$, see the main text for details. Data for various $\phi$ (ULC2 sample, shutter activated).

\clearpage
\section*{Comparison of single \textit{vs} double KWW fits to the intensity autocorrelation functions}

\begin{figure}[htbp!]
    \centering
    \includegraphics[width=0.8\textwidth]{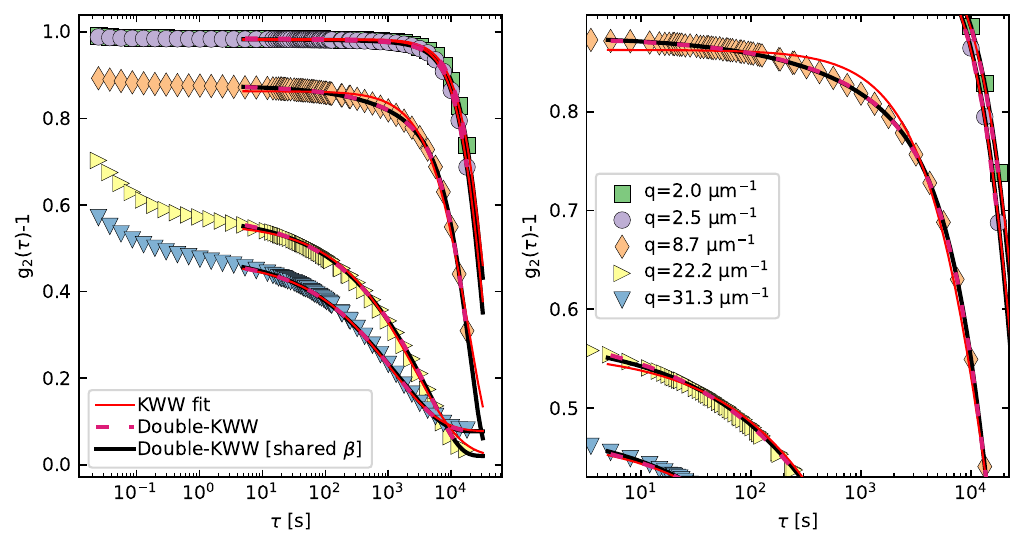}
    \caption{Intensity correlation function at various $q$, for a sample at $\varphi=1.13$, together with various fits. \textit{Left:} Best fits with a single KWW function (red continuous line), a double KWW function with $\beta$ parameters individually optimized for each $q$ (Eq. 7 of the main text, dashed pink line), and a double KWW function with $\beta$ parameters shared among data at different $q$'s (black line). \textit{Right:} Zoom of the same data as in the left panel, showing the improvement in the fitting of the experimental data with the model proposed in the main text.}
    \label{fig:fit_comparison}
\end{figure}

Figure~\ref{fig:fit_comparison} compares the quality of single- \textit{vs} double-KWW fits to $g_2-1$.

\end{document}